\documentclass[english]{article}
\usepackage[T1]{fontenc}
\usepackage[latin9]{inputenc}
\usepackage{geometry}
\usepackage[active]{srcltx}
\usepackage{color}
\usepackage{float}
\usepackage{amsmath}
\usepackage{amssymb}
\usepackage{graphicx}
\usepackage{setspace}
\usepackage{esint}

\makeatletter

\providecommand{\tabularnewline}{\\}

\makeatother

\usepackage{babel}
\begin{document}

\title{Continuum and discrete models for unbalanced woven fabrics}

\author{Angela Madeo\thanks{angela.madeo@insa-lyon.fr, LGCIE, Université de Lyon, INSA, 20 avenue
Albert Einstein, 69621, Villeurbanne cedex, France} \thanks{Corresponding author}, Gabriele Barbagallo\thanks{gabriele.barbagallo@insa-lyon.fr, LaMCoS, Université de Lyon, INSA,
20 avenue Albert Einstein, 69621, Villeurbanne cedex}, Marco Valerio D'Agostino\thanks{marco-valerio.dagostino@insa-lyon.fr, LaMCoS, Université de Lyon,
INSA, 20 avenue Albert Einstein, 69621, Villeurbanne cedex} and Philippe Boisse\thanks{philippe.boisse@insa-lyon.fr, LaMCoS, Université de Lyon, INSA, 20
avenue Albert Einstein, 69621, Villeurbanne cedex} }

\maketitle
\begin{abstract}

The classical models used for describing the mechanical behavior of
woven fabrics do not fully account for the whole set of phenomena
that occur during the testing of such materials. This lack of precision
is mainly due to the absence of energy terms related to the microstructural
properties of the fabric and, in particular, to the bending stiffness
of the yarns. The importance of the bending stiffness on the overall
mechanical behavior of woven reinforcements, if already essential
for the complete description of balanced fabrics, becomes even more
important in the case of unbalanced ones. In this paper it is shown
that the unbalance in the bending stiffnesses of the warp and weft
yarns produces macroscopic effects that are extremely visible: we
mention, for example, the asymmetric S-shape of a woven interlock
subjected to a Bias Extension Test (BET).

We propose to introduce a constrained micromorphic model and, simultaneously,
a discrete model that are both able to account for i) the angle variation
between warp and weft tows, ii) the unbalance in the bending stiffness
of the yarns and iii) the relative slipping of the tows. 

The introduced constrained micromorphic model is rigorously framed
in the spirit of the Principle of Virtual Work for the study of the
equilibrium of continuum bodies. A suitable constraint is introduced
in such micromorphic model by means of Lagrange multipliers in the
strain energy density and the resulting constrained model is seen
to tend to a particular second gradient one. The main advantage of
using such constrained micromorphic model is that the kinematical
and traction boundary conditions that can be imposed on some sub-portions
of the boundary of the considered body take a natural and unique meaning.

The discrete model is set up by opportunely interconnecting Euler-Bernoulli
beams with different bending stiffnesses in the two directions by
means of rotational and translational elastic springs. The main advantage
of such discrete model is that the slipping of the tows is described
in a rather realistic way. Suitable numerical simulations are presented
for both the continuum and the discrete models and a comparison between
the simulations and the experimental results is made showing a definitely
good agreement.

\end{abstract}

\section*{Introduction}

For decades textile composites made of woven fabrics have been successfully
employed in aircraft and automobile engineering and they are gaining
an even increased interest due to their excellent mechanical properties
such as a very high specific-strength and excellent formability properties.
Fibrous composite reinforcements present improved characteristics,
namely high specific stiffness and strength, good deformability, dimensional
stability, low thermal expansion, good corrosion resistance and many
others. Among the quoted characteristics, the good deformability is
what makes these materials perfect to be formed in various shapes
with limited expenses. On the other hand, some complex behaviors of
the woven fabrics, for instance the onset of wrinkling and slippage,
limit the admissible deformations during the stamping operations and
can render the modeling of such materials difficult to be achieved.
Due to the complexity of the micro-macro behavior of fibrous composite
reinforcements, the need of a comprehensive model for the prediction
of the mechanical response of such materials during the forming represents
a real scientific challenge (see e.g. \cite{Boisse shaping}).

\begin{figure}[h]
\begin{centering}
\includegraphics[width=14cm]{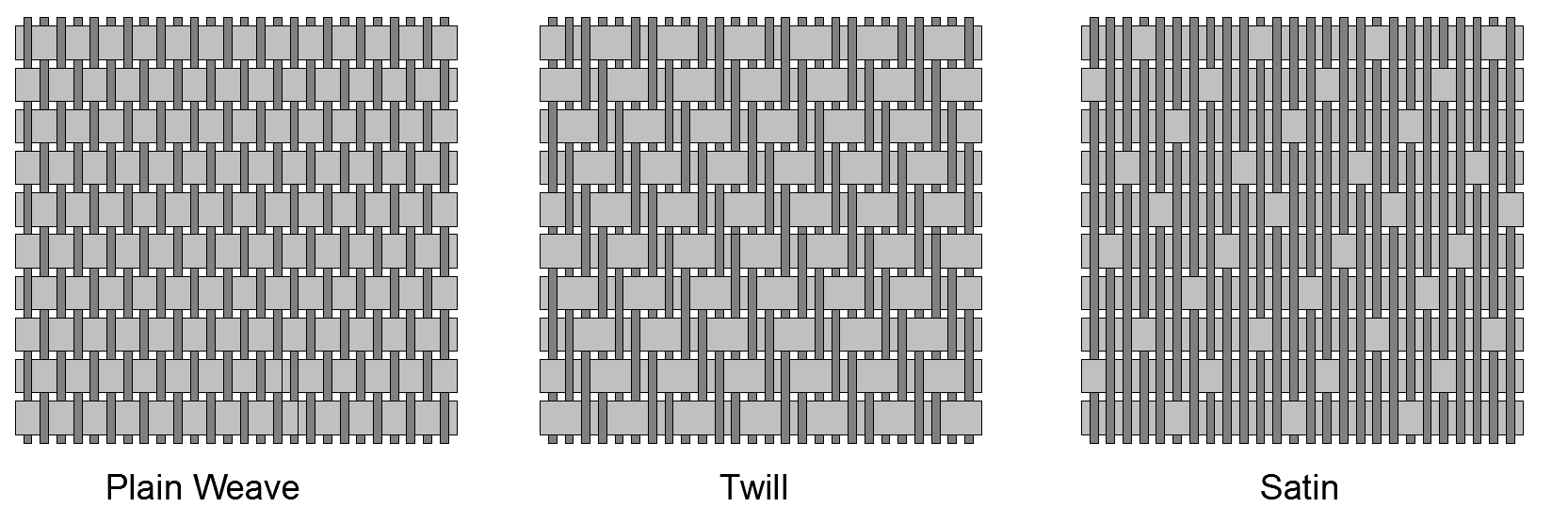}
\par\end{centering}

\protect\caption{Schemes of weaving for unbalanced fibrous composite reinforcements\label{fig:FRC-weaving}}
\end{figure}

The structure of the fabric is characterized by two main directions
of woven tows (warp and weft) and, therefore, in those directions
the fabric has a very high extensional rigidity. The way in which
these fabrics are weaved varies according to the different production
methods and it is therefore possible to observe various schemes of
weaving patterns such as those shown in figure \ref{fig:FRC-weaving}
for unbalanced fabrics. Each of these schemes leads to a different
type of composite reinforcements and thus to different mechanical
properties. Furthermore, the warp and weft of a fabric can be either
balanced (with the same properties) or unbalanced (the warp and weft
present different characteristics due, for example, to a different
number of constituting fibers). Such unbalanced fabrics may be of
use, for example, in all those engineering applications that require
a material that has to be stressed in a main direction. In such situations
the use of an unbalanced fabric conveys a real advantage in terms
of material response and therefore an economic advantage as well. 

In order to produce the fabrics, the yarns (that are themselves composed
of thousands of small carbon fibers) are weaved together forming a
very complex texture. In such a structure it is no wonder that the
interaction between the yarns and their behavior at a mesoscale play
a fundamental role on the overall response of the material. One of
the main features that determines the properties of the fabric is
the friction between the yarns that both prevents the slipping and
generates the shear rigidity of the fabric. This shear stiffness fundamentally
determines the behavior of the fabric being usually orders of magnitudes
lower than the elongation stiffness of the yarns. The shear angle
variation between warp and weft can hence be thought to be the principal
deformation mode of fibrous composite reinforcements. However, the
shear stiffness of the fabric and the elongation stiffness of the
yarns are not enough to fully describe the response of such materials.
The yarns, possess an in-plane bending stiffness that determines some
peculiar behaviors of the material at the macroscopic scale (see e.g.
\cite{dell_Steigmann,MadeoBias}). Particularly, in the case of an
unbalanced fabric the difference in the thickness of the two families
of yarns leads to peculiar responses. 

Notwithstanding the importance of a complete and effective modeling
of fibrous composite reinforcements, the most common models for the
description of the mechanical behavior of such materials fail to describe
comprehensively their response. 

\medskip{}

In this paper we propose to use a micromorphic model to describe the
mechanical behavior of unbalanced fibrous composite reinforcements
with an application to the Bias Extension Test, which is a very well
known mechanical test in the field of composite materials manufacturing
(see e.g. \cite{Cao Phi,Harrison,Lee e Phil,Peng}).

We try to limit at most the complexity of such model by introducing
a unique kinematical scalar field $\varphi$ in addition to the classical
macroscopic displacement $u$. As usual in micromorphic models, the
strain energy density of the considered material is supposed to depend
on the first gradients of both $u$ and $\varphi$. 

In order to describe at best the physics of the problem it is useful
to interpret the variable $\varphi$ as the angle variation between
warp and weft, so that a dependence of the strain energy density on
$\nabla\varphi$ actually accounts for the bending stiffness of the
yarns. Indeed, it is possible to understand that when varying the
angle from one value to a different one, this cannot be done too sharply,
but a smooth gradient of the angle variation must occur that is physically
associated to the bending of the tows.

As a matter of fact, the angle variation between the two material
directions $m_{1}$ and $m_{2}$ can be directly associated to the
gradient of the displacement field (such angle variation is known
to be the invariant $i_{8}=\left\langle m_{1},C\cdot m_{2}\right\rangle $,
of the Cauchy-Green strain tensor $C$). In order to let the micromorphic
variable $\varphi$ tend to the angle variation $i_{8}$, we decide
to use suitable Lagrange multipliers in the micromorphic strain energy
density. The resulting constrained micromorphic model can be actually
interpreted as a particular second gradient model since the first
gradient of $\varphi$ tends to the gradient of $i_{8}$ which is
itself a function of the gradient of $u$. 

As we will show in detail in the body of the paper, the proposed constrained
micromorphic model is able to account for

\begin{itemize}

\item the basic mechanism of angle variation between warp and weft
\item different bending rigidities in the warp and weft directions
\item the decrimping and eventual relative slipping of the yarns.

\end{itemize}
When applying such constrained micromorphic model to the description
of the Bias Extension Test (BET) on an unbalanced fabric, we are able
to recover the characteristic macroscopic asymmetric S-shape of the
specimen (essentially due to the two different bending stiffnesses),
as well as the deformation patterns of warp and weft tows at the mesoscopic
scale. More precisely, for what concerns the slipping of the yarns
which is experimentally observed, it is accounted for in our continuum
model by means of the introduction of ``equivalent elongations''
in the two material directions. 

\medskip{}

The reader may think that the fact of passing through a constrained
micromorphic model to formulate a specific second gradient one is
an artificial procedure that provides additional complexity to the
modeling of the considered materials. Nevertheless, we will show that
the fact of starting from a micromorphic model and to subsequently
constrain it in order to tend to a second gradient one, is a more
natural way to clarify the physical meaning of the considered problem.
In particular, the fact of considering a micromorphic model avoids
any confusion concerning the boundary conditions that can be imposed
in the considered physical problem. We will see that we will be naturally
led to impose kinematical boundary conditions that immediately take
the precise physical meaning of imposing the displacement and the
angle between warp and weft on given subsets of the boundary. Similarly,
there will be a unique way to impose forces and double forces at the
boundary of the considered micromorphic medium where tractions are
assigned, and the introduced ``force'' will be seen to be directly
related to the force which is measured by the employed testing machine. 

On the other hand, if we would have started directly from a second
gradient model the way of imposing physical boundary conditions would
have been much more complicated. In fact, it is known (see e.g. \cite{incomplete,Mindlin,Mindlin1})
that there is no a unique way to define forces and double forces in
second gradient theories, but multiple combinations of such contact
actions may be prescribed on the boundary which are all equally legitimate
but which give rise to different boundary value problems. 

It is our hope to convince the reader that the physics of the boundary
conditions which have to be imposed to model the BET is naturally
suggested by the use of the used constrained micromorphic model.

\medskip{}

In a second time, we introduce a discrete model for the description
of the Bias Extension Test on unbalanced fabrics. To do so, we use
long Euler-Bernoulli beams with different bending stiffnesses in the
two directions, which are suitably interconnected by rotational and
translational springs in such a way that the main characteristics
of the experimental Bias Extension Test are described. In particular,
we are able to recover in a realistic way both the macroscopic asymmetric
shape of the specimen as well as the pattern of the yarns at the mesoscopic
scale which include shear strain, bending and slipping. Since the
yarns constituting the specimen never experience compression during
the BET, we can sensibly affirm that the introduced model of interconnected
Euler-Bernouilli beams actually describe, at least qualitatively,
the overall behavior of the reinforcement.

Even though the proposed discrete model only involve elastic elements
(beams and springs), it is able to catch in a quite realistic way
the main features of the deformation of unbalanced fabrics. A point
that could be improved in this sense is to introduce some dispersion
in the model (friction), following e.g. the methods presented in \cite{Gatouillat,David4},
in order to account for the phenomenon which is experimentally observed
and which suggests that, after unloading, the specimen does not perfectly
return to its initial undeformed shape.

The results obtained via the discrete model allow us to comfort those
obtained via the constrained micromorphic continuum model and they
additionally allow to bring more light on the mechanisms of slipping
that occur during the BET on unbalanced fabrics.

\medskip{}

The paper is organized as follows
\begin{itemize}
 
\item In section 1 we set-up the equilibrium problem for both Cauchy and
micromorphic continua. To do so, we introduce suitable spaces of configurations
and of admissible variations whose structure is directly related to
the kinematical boundary conditions that are imposed in the considered
problem. We hence formulate the equilibrium problem for the considered
(Cauchy and Micromorphic) continua by means of the use of the Principle
of Virtual Work. Finally, we treat in more detail the problem of the
equilibrium of a first gradient continuum by obtaining the irreducible
form of the work of internal actions for such a continuum. This irreducible
form allows to individuate the type of boundary contact actions which
can be introduced in Cauchy media, namely ``forces'' per unit area.
Finally, we explicitly show a method to calculate external contact
actions by means of the use of the Principle of Virtual Work: such
method is based on a wise choice of particular test functions that
allow to isolate forces and double forces on some specific parts of
the boundary.
\item In section 2 we study the equilibrium of an unbalanced 2D fabric subjected
to a Bias Extension Test. This study is carried out following three
subsequent steps. First of all the physical problem (BET on unbalanced
fabric) is precisely described with particular attention to the description
of the imposed kinematical and boundary conditions (see e.g. \cite{Cao Phi,Phil1,Gatouillat,Lee e Phil}).
Secondly, a discussion concerning the most appropriate continuum model
which is needed to describe the physical phenomenon of interest is
carried out. We propose a constrained micromorphic model to accomplish
this task (see also \cite{MadeoBias}). Such model is able to account
for i) the angle variation between warp and weft yarns, ii) the unbalanced
bending stiffness of the two families of fibers and iii) the relative
slipping of the yarns by means of the introduction of ``equivalent
elongations''. Last but not least the introduced constrained micromorphic
model is seen to be able to introduce in a natural way the boundary
conditions which are peculiar of the BET. Finally, we present the
numerical simulations of the proposed constrained micromorphic continuum
model to show that it is able to satisfactorily describe the experimental
evidences.
\item In section 3 we introduce the discrete model for the description of
the BET by means of the use of Euler-Bernoulli beams suitably interconnected
by rotational and translational elastic springs. We show that the
obtained results fit well the available experimental evidences both
for what concerns the macroscopic and microscopic deformation patterns.
In particular, the phenomenon of relative slipping of the yarns is
unveiled in a rather precise way so allowing a wise interpretation
of the equivalent elongations introduced in the continuum micromorphic
model.   

\end{itemize}
 
We explicitly remark that section 1 contains an introductory theoretical
treatise which is convenient to frame the considered mechanical problem
in the framework of the Principle of Virtual Work. We believe that
such a discussion is indeed very useful for those readers who want
to make a clear connection between a neat mathematical formulation
of the Principle of Virtual Work and the description of the mechanical
phenomena that such principle can provide. Actually, an intelligible
and self-consistent disquisition ranging from the mathematical setting
to the mechanical interpretation of the Principle of Virtual Work
is difficult to be found in academic articles, so that we believe
that section 1 represents an added value for the present paper. Some
of the introduced notations are voluntarily lightened with respect
to those presented in classical more mathematical books (see e.g.
\cite{Marsden}) in order to make easier the connection between the
mathematical and the mechanical aspects of the same problem. The reader
who is not interested in establishing such connection can skip this
opening section, directly passing to the mechanical setting-up of
the considered problem. On the other hand, the reader who is interested
in a more precise and general mathematical formulation of the problem
can refer e.g. to \cite{Marsden}.

\section{\label{sec:The-Principle-of}The Principle of Virtual Work and the
equilibrium of Cauchy and micromorphic continua}

In what follows we will call \textit{\textcolor{black}{Cauchy continuum
body}} a set of material particles occupying the volume $B$ in its
reference configuration and whose motion is described by means of
a suitably regular, kinematical field $u:B\rightarrow\mathbb{R}^{3}$
which we call the \textit{\textcolor{black}{displacement field}} of
the considered body. We denote by\footnote{Here and in the sequel we denote by $T\mathbb{R}^{3}$ the tangent
bundle to the manifold $\mathbb{R}^{3}$. We recall that the virtual
variation $\delta u$ of a displacement field $u$ has the structure
$\delta u(X)=(u(X),\delta w(X))$, where $\delta w(X)$ is a vector
attached to the Eulerian point $u(X)$.} $\delta u:B\rightarrow T\mathbb{R}^{3}$ the \textit{\textcolor{black}{virtual
displacement field}} associated to the considered body. Finally, we
will denote by $\partial B$ the boundary of $B$.

Generalizing this definition of Cauchy continuum, we will introduce
a simple \textit{\textcolor{black}{micromorphic continuum body}} by
complementing the previously defined Cauchy continuum with a supplementary,
suitably regular, scalar kinematical field $\varphi:B\rightarrow\mathbb{R}$,
that we generally call \textit{\textcolor{black}{micro-motion}}. We
denote by\footnote{Here and in the sequel we denote by $T\mathbb{R}$ the tangent bundle
to the manifold $\mathbb{R}$. We recall that the virtual variation
$\delta\varphi$ of a micro-strain field $\varphi$ has the structure
$\delta\varphi(X)=(\varphi(X),\delta\psi(X))$, where $\delta\psi(X)$
is a vector attached to the image point $\varphi(X)$.} $\delta\varphi:B\rightarrow T\mathbb{R}$ the virtual variations
of the kinematical field $\varphi$. We remark that, in the spirit
of Mindlin \cite{Mindlin} and Eringen \cite{EringenBook}, such supplementary
kinematical field represents the motion of a microstructure which
is embedded in the considered body. Such micro-motion is, in principle,
completely independent of the macroscopic motion of the matrix. Nevertheless,
in some cases of physical interest, as the one which will be analyzed
in this paper, it is worth to relate such micro-descriptors to the
derivatives of the macroscopic displacement field. As we will show
later on in much more detail, this can be done by constraining the
introduced micromorphic model with suitable Lagrange multipliers to
be added in the strain energy density, or equivalently, by introducing
a penalty term in the energy itself. We want to stress the fact that
we prefer to keep a more general micromorphic model and to subsequently
constrain its strain energy density in order to let it tend to a second
gradient one. This choice is preferable if one wants to interpret
in a unique way the external actions of the considered continuum.
In fact, (see e.g. \cite{incomplete,Mindlin,Mindlin1}) if one starts
directly from a second gradient energy, the interpretation of boundary
contact actions, namely forces and double-forces, depends on the type
of manipulation which is done on the work of internal actions by means
of procedures of integration by parts. More particularly, if one decides
to stop at a given level of integration by parts, or to continue further
to the subsequent level, the definition of force and double force
is not the same\footnote{\begin{singlespace}
We need to mention the fact that no common agreement is currently
available concerning the choice of different but equally legitimate
sets of boundary conditions deriving to different levels of integration
by parts. Nevertheless, this is what is found e.g. in \cite{incomplete,Mindlin,Mindlin1}
and we tend to adopt this viewpoint in the recent times.\end{singlespace}
}. When considering micromorphic continua in which only first gradient
of the introduced kinematical fields appear in the strain energy density,
only one level of integration by parts is possible and then the boundary
contact actions are uniquely defined and take immediate physical meaning
when framed in the considered physical problem.

\medskip{}

The equilibrium of a Cauchy continuum body subjected to given boundary
conditions can be studied by means of the Principle of Virtual Work.
Such fundamental principle of Mechanics states that a body, subjected
to specific external actions, is in equilibrium if the work of internal
actions is balanced by the work of external actions. In formulas,
we say that a displacement field $u^{*}$ is an equilibrium configuration
if \footnote{We remark that, depending on the conventions which are used for the
signs in the definition of the work of internal and external actions,
slightly different versions of the Principle of Virtual Work can be
found in the literature.}
\begin{equation}
\mathcal{P}^{int}\left(u^{*},\delta u\right)+\mathcal{P}^{ext}\left(u^{*},\delta u\right)=0,\label{eq:PVP}
\end{equation}
for any compatible $\delta u$. In most cases, the external and internal
works can be seen as the first variation of suitable functionals $\mathcal{A}^{int}\left(u\right):Q\rightarrow\mathbb{R}$
and $\mathcal{A}^{ext}\left(u\right):Q\rightarrow\mathbb{R}$, so
that the Principle of Virtual Work (\ref{eq:PVP}) actually implies
the minimization of a functional $\mathcal{A}:=\mathcal{A}^{int}+\mathcal{A}^{ext}.$
More specifically, we can write
\begin{equation}
\mathcal{P}^{int}\left(u,\delta u\right)+\mathcal{P}^{ext}\left(u,\delta u\right)=\delta\mathcal{A}(u,\delta u):=\lim_{t\rightarrow0^{+}}\frac{\mathcal{A}\left(u+t\,\delta u\right)-\mathcal{A}(u)}{t},\label{eq:PVP-1}
\end{equation}
where we denoted by $\delta\mathcal{A}$ the first variation of the
functional $\mathcal{A}$, where $u\in Q$, $\delta u\in T_{u}$ and
the sets $Q$ and $T_{u}$ will be defined in more detail later on. 

\medskip{}

Suitably generalizing the definitions given above for first gradient
continua, the Principle of Virtual Work can be reformulated for the
introduced micromorphic continuum by saying that a couple $(u^{*},\varphi^{*})$
is of equilibrium if 
\begin{equation}
\mathcal{P}^{int}\left(u^{*},\varphi^{*},\delta u,\delta\varphi\right)+\mathcal{P}^{ext}\left(u^{*},\varphi^{*},\delta u,\delta\varphi\right)=0,\label{eq:PVP-2}
\end{equation}
for any compatible $\left(\delta u,\delta\varphi\right)$.

Again, the external and internal works can be seen as the first variation
of suitable functionals $\mathcal{A}^{int}\left(u,\varphi\right):Q\times D\rightarrow\mathbb{R}$
and $\mathcal{A}^{ext}\left(u,\varphi\right):Q\times D\rightarrow\mathbb{R}$,
so that the Principle of Virtual Work (\ref{eq:PVP-2}) actually implies
the minimization of a functional $\mathcal{A}:=\mathcal{A}^{int}+\mathcal{A}^{ext}.$
More specifically, we can write 
\begin{equation}
\mathcal{P}^{int}\left(u,\varphi,\delta u,\delta\varphi\right)+\mathcal{P}^{ext}\left(u,\varphi,\delta u,\delta\varphi\right)=\delta\mathcal{A}(u,\varphi,\delta u,\delta\varphi):=\lim_{t\rightarrow0^{+}}\frac{\mathcal{A}\left(u+t\,\delta u,\varphi+t\,\delta\varphi\right)-\mathcal{A}(u,\varphi)}{t},\label{eq:PVP-1-1}
\end{equation}
where we denoted again by $\delta\mathcal{A}$ the first variation
of the functional $\mathcal{A}$, where $\left(u,\varphi\right)\in Q\times D$,
$\left(\delta u,\delta\varphi\right)\in T_{u}\times T_{\varphi}$
and the sets $Q$, $D$, $T_{u}$ and $T_{\varphi}$ will be defined
in more detail later on. 

\medskip{}

Suitable generalizations of the Principle of Virtual Work can be introduced
in order to account for inertial and dissipative effects, but, since
we deal in this paper only with static problems, we refrain here to
present such more complex framework. 

\medskip{}

The most fundamental questions which have to be confronted to properly
set up a mechanical theory by means of the Principle of Virtual Work
is to establish:
  
\begin{itemize}
 
\item the constitutive form of the\textit{ work of internal actions} in
terms of the displacement and, eventually, of the micro-descriptor
(such constitutive choice is related to the intrinsic nature of the
medium that one wants to study),
\item the expression of the \textit{work of external actions}, which allows
to establish how the external world acts on the considered medium
and to define the concept of force, double force, or other more complex
interactions.  

\end{itemize}
 
\medskip{}

As a matter of fact, we can imagine to act on the boundary of the
considered body by imposing either
  
\begin{itemize}
 
\item kinematical (or essential or geometric) boundary conditions: the displacement
and/or eventually the micro-descriptor are assigned on some portion
$\Sigma_{K}$ of the boundary $\partial B$,
\item traction (or natural) boundary conditions: forces and/or, eventually,
other more complex external interactions are assigned on some portion
$\Sigma_{T}$ of the boundary $\partial B$.  

\end{itemize}
 
In order to be more general, we can introduce 
  
\begin{itemize}
 
\item The surface $\Sigma_{K_{1}}$ on which displacement is assigned and
the surface $\Sigma_{K_{2}}$ on which we can eventually assign the
micro-descriptor $\varphi$. Clearly, $\Sigma_{K}=\Sigma_{K_{1}}\bigcup\Sigma_{K_{2}}$,
and the two sets $\Sigma_{K_{1}}$ and $\Sigma_{K_{2}}$ may eventually
partially or totally overlap depending on the considered physical
problem. 
\item The surfaces $\Sigma_{T_{1}}$ on which we assign forces and $\Sigma_{T_{2}}$
on which we can assign more complex interactions and that can as well
partially or totally overlap depending on the physical problem in
study. We also have $\Sigma_{T}=\Sigma_{T_{1}}\bigcup\Sigma_{T_{2}}$.   

\end{itemize}
 
If, in the considered physical problems (this may eventually arrive
in micromorphic theories), mixed kinematical-traction boundary conditions
are applied, also the sets $\Sigma_{K_{1}}$ and $\Sigma_{T_{2}}$
and/or $\Sigma_{K_{2}}$ and $\Sigma_{T_{1}}$ may eventually have
non-vanishing intersection.

We also explicitly mention that, in order to be able to recover all
the possible external interactions, each of the introduced sets $\Sigma_{K_{1}}$,
$\Sigma_{K_{2}}$, $\Sigma_{T_{1}}$ and $\Sigma_{T_{2}}$ may collapse
in the empty set or can cover the whole boundary $\partial B$. Finally,
we remark that, by definition, we set $\Sigma_{K}\bigcup\Sigma_{T}=\partial B$.
Finally, we also mention that we will neglect external volume forces
in the following treatment.

\subsection{\label{sub:Space-of-configurations}Space of configurations and spaces
of admissible variations}

Depending on the intrinsic nature of the considered body (first gradient
or micromorphic), the expressions for the work of internal and external
actions take specific forms which will be better specified later on.

Independently of the specific form taken by the internal and external
work, we want to underline here that the problem of finding the equilibrium
configuration of a given continuum (Cauchy or micromorphic) subjected
to specific boundary conditions reduces to the problem of finding,
in suitable sets, the kinematical fields which satisfy the Principle
of Virtual Work ((\ref{eq:PVP}) or (\ref{eq:PVP-2})) for any virtual
admissible variations.

\medskip{}

We start by introducing the equilibrium problem for a Cauchy continuum,
defining a suitable set $Q$, called \textit{\textcolor{black}{space
of configurations}} of the considered medium which contains information
about the kinematical constraints which must be verified by the displacement
field. More precisely, we define the space of configurations for a
Cauchy continuum as\footnote{We remark that the set $Q$ should also contain informations concerning
the desired regularity on $u$, but we limit ourselves here to talk
about ``suitably regular'' functions.} 
\begin{equation}
Q=\left\{ u\ \mid\ u=\bar{u}\ \ \text{on\ }\ \Sigma_{K_{1}}\right\} ,\label{eq:SC1st}
\end{equation}
where $\bar{u}$ is a suitably assigned function. Roughly speaking,
the set $Q$ represents the set in which we look for the solution
of our minimization problem and contains only those displacement fields
which satisfy the imposed kinematical boundary conditions.

On the other hand, we define the set of admissible variations as
\begin{equation}
T_{u}=\left\{ \delta u\ \mid u+\delta u\in Q\right\} .\label{eq:VA1st}
\end{equation}
We explicitly remark that, being $u=\bar{u}$ on $\Sigma_{K_{1}}$,
in order to have $\delta u$ belonging to the set of admissible variations,
one must have that $\bar{u}+\delta u=\bar{u}$ on $\Sigma_{K_{1}}$.
This clearly implies that $\delta u=0$ on $\Sigma_{K_{1}}$, and
hence the set of admissible variations takes the form $T_{u}=\left\{ \delta u\ \mid\delta u=0\ \ \text{on\ }\ \Sigma_{K_{1}}\right\} $.With
the introduced notations, we can formulate the equilibrium problem
for a Cauchy continuum as:

\textit{\textcolor{black}{Find $u^{*}\in Q$ such that $\mathcal{P}^{int}(u^{*},\delta u)+\mathcal{P}^{ext}(u^{*},\delta u)=0\quad$
for any $\delta u\in T_{u^{*}}$.}}

\medskip{}
The setting-up of the equilibrium problem for a micromorphic continuum
can be obtained, suitably generalizing what done for Cauchy continua.
In particular, a supplementary set $D$ must be defined to specify
the space of configurations for such continuum\footnote{We remark that the set $D$ should also contain informations concerning
the desired regularity on $\varphi$, but we limit ourselves here
to talk about ``suitably regular'' functions.}: 
\[
D=\left\{ \varphi\mid\ \varphi=\bar{\varphi}\ \ \text{on}\ \ \Sigma_{K_{2}}\right\} ,
\]
where $\bar{\varphi}$ is a suitably assigned function. Roughly speaking,
the set $D$ represents the set in which we look for the solution
for the micro-motion of our minimization problem and contains only
those micro-motions which satisfy the imposed kinematical boundary
conditions.

Moreover, we define a supplementary set of admissible variations as
\[
T_{\varphi}=\left\{ \delta\varphi\ \mid\varphi+\delta\varphi\in D\right\} .
\]
With the introduced notations, we can formulate the equilibrium problem
for a micromorphic continuum as:

\textit{\textcolor{black}{Find $\left(u^{*},\varphi^{*}\right)\in Q\times D$
such that $\mathcal{P}^{int}(u^{*},\varphi^{*},\delta u,\delta\varphi)+\mathcal{P}^{ext}(u^{*},\varphi^{*},\delta u,\delta\varphi)=0\quad$
for any $\left(\delta u,\delta\varphi\right)\in T_{u^{*}}\times T_{\varphi^{*}}$.}}

\subsection{\label{sub:Equil_1st_Grad}The example of the equilibrium of a first
gradient continuum}

As it has been previously pointed out, in order to establish the equilibrium
problem for a given continuum body subjected to specific external
interactions, the expressions of both the work of internal and external
actions must be specified.

For a first gradient continuum the work of internal actions is defined
through the definition of the action functional $\mathcal{A}^{int}$
which can be introduced in the static case as\footnote{Classically, in the dynamic case the internal action functional is
defined as the space-time integral of the Lagrangian density $\text{\ensuremath{\mathcal{L}}}=T-W$,
where $T$ is the kinetic energy density. In the particular case of
statics, the minus sign remains after the due simplifications.}
\[
\mathcal{A}^{int}=-\int_{B}W\left(\nabla u\right)\,dv,
\]
where $W$ is the strain energy density which, in a first gradient
model, constitutively depend only on the first gradient of displacement.

In the case of first gradient theories we can hence recognize that
the work of internal actions can be written as \footnote{The operator $\mathrm{Div}$ stands for the classical divergence operator.
Being $A$ a tensor field of any order $n>0$, we define its divergence
as the $n-1$ tensor $\left(\mathrm{Div}A\right)_{i_{1},\dots i_{n-1}}=A_{i_{1},\dots i_{n},i_{n}}$.
Finally $\left\langle a,b\right\rangle =a_{i_{1},\dots i_{n}\text{ }}b_{i_{1},\dots i_{n}\text{ }}$
indicates the scalar product between two tensors of any order $n\geq1$
and the Einstein convention of sum over repeated indices is used.} 
\begin{align}
\mathcal{P}^{int}(u,\delta u) & =-\int_{B}\delta\,W\left(\nabla u\right)\,dv=-\int_{B}\left\langle \frac{\partial W}{\partial\nabla u}\ ,\ \nabla\delta u\right\rangle \,dv\nonumber \\
\label{eq:Irreducible1st}\\
 & =\int_{B}\left\langle \mathrm{Div}\left(\frac{\partial W}{\partial\nabla u}\right)\ ,\ \delta u\right\rangle \,dv-\int_{\partial B}\left\langle \frac{\partial W}{\partial\nabla u}\cdot n\ ,\ \delta u\right\rangle \,ds,\nonumber 
\end{align}
where to obtain the last identity the divergence theorem has been
used. Equation (\ref{eq:Irreducible1st}) furnishes the \textit{irreducible
expression of the work of internal actions} for a first gradient continuum.
This means that no more integrations by parts can be performed to
ulteriorly manipulate this expression of $\mathcal{P}^{int}$. It
can be remarked from such irreducible form of the internal work that,
as far as the boundary $\partial B$ is concerned, only quantities
expending work on the virtual displacement $\delta u$ (i.e. forces)
can be recognized. It is for this reason that, based on the validity
of the Principle of Virtual Work, we can affirm that the only boundary
external actions that can be sustained by a first gradient continuum
are forces per unit area, i.e. external actions expending work on
$\delta u$. These observations are at the origin of the introduction
of the work of external actions for first gradient continua in the
form\footnote{We recall once again that, to the sake of conciseness, we suppose
in this paper that bulk external actions are vanishing.}
\begin{equation}
\mathcal{P}^{ext}(u,\delta u)=\int_{\Sigma_{T_{1}}}\left\langle f,\delta u\right\rangle ds,\label{eq:Pext_1st}
\end{equation}
where $f:B\rightarrow\mathbb{R}^{3}$ is a suitable function assigned
on $\Sigma_{T_{1}}$. 

Once assigned the specific form for the work of internal and external
actions, the equilibrium problem for a first gradient continuum can
be hence reformulated as follows:

\textit{\textcolor{black}{Find $u^{*}\in Q$ such that $\mathcal{P}^{int}(u^{*},\delta u)+\mathcal{P}^{ext}(u^{*},\delta u)=0\quad$
for any $\delta u\in T_{u^{*}}$,}}

where now $\mathcal{P}^{int}$, $\mathcal{P}^{ext}$, $Q$ and $T_{u^{*}}$
are given by (\ref{eq:Irreducible1st}), (\ref{eq:Pext_1st}), (\ref{eq:SC1st})
and (\ref{eq:VA1st}) respectively.

\subsubsection{\label{sub:Reaction_Force}Evaluation of the reaction force corresponding
to an imposed boundary displacement. }

Once that the solution $u^{*}$ of the equilibrium problem for a first
gradient continuum has been found following the steps presented in
subsection \ref{sub:Equil_1st_Grad}, it may be interesting to know
which is the reaction force acting on $\Sigma_{K_{1}}$ that balances
the displacement $\bar{u}$ which has been imposed on the surface
$\Sigma_{K_{1}}$ itself. In other words, we are looking for the force
that one should apply on the portion of the boundary $\Sigma_{K_{1}}$
in order to produce the displacement $\bar{u}$ on $\Sigma_{K_{1}}$
and the displacement field $u^{*}$ within the considered body. In
order to answer to this question, it is possible to pass again through
the Principle of Virtual Work, but imagining now that a work of external
forces must be introduced also on the portion of the boundary $\Sigma_{K_{1}}$.
We hence re-define the external work (\ref{eq:Pext_1st}) by adding
an additional term as follows 
\begin{equation}
\mathcal{P}^{ext}(u,\delta u)=\int_{\Sigma_{T_{1}}}\left\langle f,\delta u\right\rangle ds+\int_{\Sigma_{K_{1}}}\left\langle f_{R},\delta u\right\rangle ds.\label{eq:Pext_1st-1}
\end{equation}
The reaction force can hence be calculated by writing the Principle
of Virtual Work as 
\begin{equation}
\int_{B}\left\langle \mathrm{Div}\left(\frac{\partial W}{\partial\nabla u^{*}}\right)\ ,\ \delta u\right\rangle \,dv-\int_{\partial B=\Sigma_{T_{1}}\bigcup\Sigma_{K_{1}}}\left\langle \frac{\partial W}{\partial\nabla u^{*}}\cdot n\ ,\ \delta u\right\rangle \,ds+\int_{\Sigma_{T_{1}}}\left\langle f,\delta u\right\rangle ds+\int_{\Sigma_{K_{1}}}\left\langle f_{R},\delta u\right\rangle ds=0\label{eq:PVP_IPP}
\end{equation}
and imposing that it must be valid for any arbitrary displacement
$\delta u$. Such identity must be satisfied, in particular, for a
virtual displacement field $\delta\bar{u}$ which is such that
  
\begin{itemize}
 
\item $\delta\bar{u}$ is constant on $\Sigma_{K_{1}}$ 
\item $\delta\bar{u}$ is continuous on $\partial B$
\item $\delta\bar{u}$ is vanishing outside $\Sigma_{K_{1}}$ except on
a suitably small region in order to preserve continuity with the imposed
displacement at the boundary.  

\end{itemize}
 
For such particular test function, the Principle of Virtual Work (\ref{eq:PVP_IPP})
implies

\[
\int_{\Sigma_{K_{1}}}\left\langle f_{R},\delta\bar{u}\right\rangle ds=\left\langle \int_{\Sigma_{K_{1}}}f_{R}\,ds\ ,\delta\bar{u}\right\rangle =\left\langle \int_{\Sigma_{K_{1}}}\frac{\partial W}{\partial\nabla u^{*}}\cdot n\,ds\ ,\ \delta\bar{u}\right\rangle .
\]
The constant test field $\delta\bar{u}$ can hence be simplified on
the two sides and, introducing the quantity $R:=\int_{\Sigma_{K}}f_{R}\,ds$,
we can finally write
\begin{equation}
R=\int_{\Sigma_{K_{1}}}\frac{\partial W}{\partial\nabla u^{*}}\cdot n\,ds.\label{eq:Reac1}
\end{equation}
We call $R$ the \textit{reaction force} associated to the imposed
displacement $\bar{u}$ on $\Sigma_{K_{1}}$. We will see that the
fact of computing the reaction forces on the basis of the procedure
shown here may be of interest for comparing the results of the performed
numerical simulations to the experimental data. In fact, when a standard
testing machine measures a ``force'' associated to an imposed displacement,
it is actually measuring a resultant force on the considered boundary.

We explicitly remark that, an equivalent way to calculate the reaction
force can be found considering the work of internal forces before
integrating by parts as given in the first equality in expression
(\ref{eq:Irreducible1st}), i.e.
\[
\mathcal{P}^{int}(u,\delta u)=-\int_{B}\left\langle \frac{\partial W}{\partial\nabla u}\ ,\ \nabla\delta u\right\rangle \,dv.
\]

With reasoning analogous to the previous ones, if we write the Principle
of Virtual Work evaluated in the equilibrium solution $u^{*}$, we
have 
\[
-\int_{B}\left\langle \frac{\partial W}{\partial\nabla u^{*}}\ ,\ \nabla\delta u\right\rangle \,dv+\int_{\Sigma_{T_{1}}}\left\langle f,\delta u\right\rangle ds+\int_{\Sigma_{K_{1}}}\left\langle f_{R},\delta u\right\rangle ds=0,
\]
which must be satisfied for any $\delta u$. If we hence choose a
particular test function $\delta\bar{u}$ such that
  
\begin{enumerate}
 
\item $\delta\bar{u}$ is constant on $\Sigma_{K_{1}}$
\item $\delta\bar{u}$ is continuous on $\partial B$
\item $\delta\bar{u}$ is an arbitrarily assigned, non-vanishing function
outside $\Sigma_{K_{1}}$  

\end{enumerate}
 
we can evaluate the reaction force by noticing that 
\begin{equation}
\left\langle R,\delta\bar{u}\right\rangle =\int_{B}\left\langle \frac{\partial W}{\partial\nabla u^{*}}\ ,\ \nabla\delta\bar{u}\right\rangle \,dv-\int_{\Sigma_{T_{1}}}\left\langle f,\delta\bar{u}\right\rangle ds.\label{eq:Reac2}
\end{equation}
We explicitly remark that Eq. (\ref{eq:Reac2}) is a scalar equation,
which means that the three components of the vector $R$ cannot be
directly calculated only using such equation. Nevertheless, if we
choose suitable test functions $\delta\bar{u}$ which are aligned
with one of the directions of the used reference system, we can arrive
to evaluate separately the components of $R$. In particular, let
$\left\{ e_{i}\right\} _{i\in\left\{ 1,2,3\right\} }$, be an orthonormal
basis with respect to which we want to evaluate the three components
$R_{i}=\left\langle R,e_{i}\right\rangle $ of the reaction force
$R$. We choose a test function $\delta\bar{u}$ which possesses all
the characteristics previously listed except that the point 1. is
replaced by 
  
\begin{itemize}
 
\item $\delta\bar{u}$ is equal to $e_{i}$ on $\Sigma_{K_{1}}$.  

\end{itemize}
 
In this case the component $R_{i}$ can be easily calculated according
to Eq. (\ref{eq:Reac2}).

Clearly, if the problem is well formulated, the value of the reaction
$R$ calculated with expression (\ref{eq:Reac1}) must coincide with
the one calculated using expression (\ref{eq:Reac2}). The second
way of evaluating the reaction force has many advantages, especially
when we want to do it numerically. Indeed, it is always more stable
to numerically evaluate a volume integral than a surface integral.
It is for this reason that, for the case of micromorphic continua
shown in the next section, we will limit ourselves to show how to
calculate reaction forces and double-forces, by passing trough the
evaluation of volume integrals.

\section{Equilibrium of a 2D unbalanced woven fabric modeled as an orthotropic
micromorphic continuum}

It has been known since the pioneering works by Piola \cite{Piola},
Cosserat \cite{Cosserat}, Midlin \cite{Mindlin}, Toupin \cite{Toupin},
Eringen \cite{EringenBook}, Green and Rivlin \cite{GreenRivlin}
and Germain \cite{Germain,Germain2} that many microstructure-related
effects in mechanical systems can be still modeled by means of continuum
theories. It is known since then that, when needed, the placement
function must be complemented by additional kinematic descriptors,
called sometimes micro-structural fields. More recently, these generalized
continuum theories have been widely developed (see e.g. \cite{NthGrad,BC2grad1,BC2grad2,BC2grad3,Hooke2grad,Aifantis,Forest Aifantis,Trianta,Forest,Forest0,Victor00,Victor2})
to describe the mechanical behavior of many complex systems, such
as e.g. porous media \cite{FdIGuarHutter,madeo porous,CoussiSciarra,sciarra},
capillary fluids \cite{Casal,de Gennes,FdI cap1,IsolaRotoli,FdI cap2},
exotic media obtained by homogenization of heterogeneous media \cite{Alibert,Pideri-Sepp,Seppecher Exotic}.
Interesting applications on wave propagation in such generalized media
has also gained attention in the recent years for the possible application
of this kind of materials to passive control of vibrations and stealth
technology (see e.g. \cite{FdIwaves,madeoWavePor,Placidi,Guyader}).

\medskip{}

In this section, we will frame the problem of determining the equilibrium
of a 2D, unbalanced, woven reinforcement modeled as a micromorphic
material, in the spirit of section \ref{sec:The-Principle-of}. With
this precise aim in mind, we will first present the particular physical
problem that we want to study, then we will propose suitable expressions
for the work of internal and external actions and we will introduce
the adapted spaces of configurations and of admissible variations.
We will finally formulate the equilibrium problem for the introduced
micromorphic continuum and we will solve it by means of suitable numerical
simulations.

\subsection{\label{sub:The-physical-problem:}The physical problem: bias extension
test on an unbalanced fabric}

We consider in this subsection the description of the physical problem
that we want to analyze and for which we will develop a suitable mechanical
second constrained micromorphic model in the next subsection. The
material that we consider here is an unbalanced fabric, i.e. a fibrous
reinforcement in which the warp and weft yarns have very different
thickness due to the fact that they are constituted by a significantly
different number of fibers. This unbalance is reflected in a different
mechanical behavior of the two orders of yarns. In order to sketch
a simplified description of the mechanical behavior of unbalanced
fabrics we can notice that
  
\begin{itemize}
 
\item the yarns can be supposed to be \textit{\textcolor{black}{quasi-inextensible}}
in both the warp and weft directions, due to the very high tensile
resistance of the constituent carbon fibers. This means that, notwithstanding
the different thickness of the yarns, we can suppose that they do
not sensibly elongate. Actually, an initial apparent elongation of
the yarns due to \textit{\textcolor{black}{decrimping}} can be eventually
observed when testing the woven material, but it is reasonable to
suppose that such apparent elongation, if present, has eventually
the same characteristics in both directions. 
\item The warp and weft being strongly unbalanced, we can infer that they
possess very \textit{\textcolor{black}{different bending stiffnesses}}.
More particularly, the thick yarns are sensible to exhibit a higher
resistance to bending than the thin ones.
\item Depending on the nature of the externally applied load and/or boundary
conditions, the yarns can experience some \textit{\textcolor{black}{relative
slipping}}. More precisely, it is possible that the contact point
between two yarns can move during the deformation of the macroscopic
piece.  

\end{itemize}
 
\begin{figure}[h]
\begin{centering}
\includegraphics[angle=90,scale=0.5]{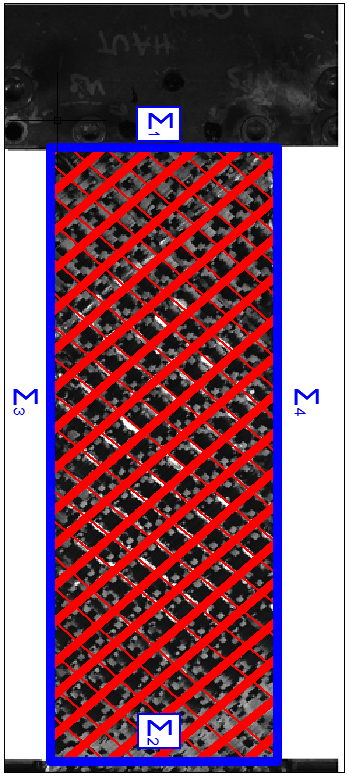}
\par\end{centering}

\protect\caption{\label{fig:Experimental_Set_up}Experimental set-up for a bias extension
test on an unbalanced fabric: it must be $\left|\Sigma_{4}\right|\geq2.5\left|\Sigma_{1}\right|$.
For the considered experimental test we have $\left|\Sigma_{1}\right|=70\,mm$
and $\left|\Sigma_{4}\right|=220\,mm$. Moreover, the specimen is
$15\:mm$ thick, but we suppose that this has no influence on the
results, i.e. no displacement neither deformation occur out of the
plane of the fabric.}
\end{figure}

We are interested in this paper to the description of a Bias Extension
Test on an unbalanced fabric of the type described above. More precisely,
the specimens object of the study are rectangular carbon fiber interlocks
(the height must at least 2.5 times longer than the basis) with unbalanced
yarns oriented at $\pm45^{\circ}$ with respect to the long side of
the specimen (see Fig. \ref{fig:Experimental_Set_up}). The two shorter
edges of the specimens are clamped in suitable devices which assure
the following kinematical boundary conditions
  
\begin{itemize}
 
\item vanishing displacement on $\Sigma_{1}$,
\item constant assigned displacement $u_{0}$ on $\Sigma_{2}$,
\item angle between warp and weft blocked at $45^{\circ}$ on $\Sigma_{1}$
and $\Sigma_{2}$ (i.e. vanishing angle variation between the two
families of yarns during the motion of the fabric).  

\end{itemize}
 
\begin{figure}[h]
\centering{}\includegraphics[width=14cm]{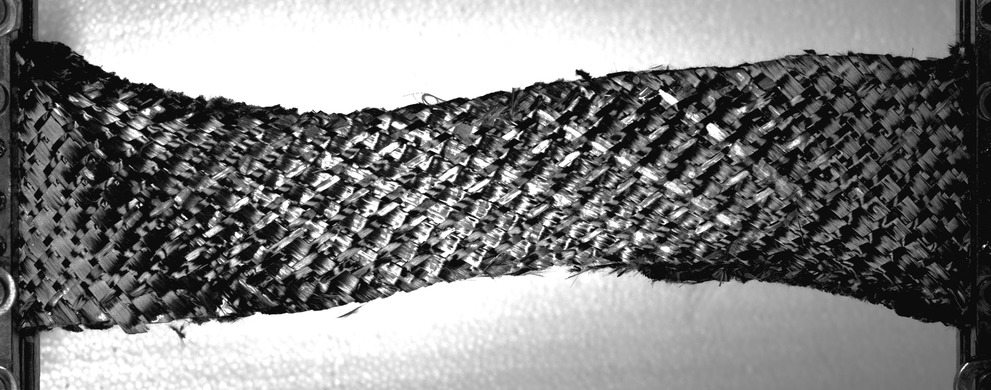}\protect\caption{\label{fig:Exp_S_Shape}Experimental deformed shape for an imposed
displacement $u_{0}=56\,mm$.}
\end{figure}

The experimental result of the performed mechanical test is a S-shaped
macroscopic deformation as the one presented in figure \ref{fig:Exp_S_Shape}.

The obtained asymmetric shape is related to the fact that the material
properties and, in particular, the bending stiffnesses of the yarns
are not the same in the two privileged material directions. Moreover,
due to the applied boundary conditions, some non-negligible slipping
of the yarns is also observed, above all for what concerns the central
part of the specimen in which yarns with two free ends are located.

\begin{figure}[h]
\centering{}\includegraphics[width=14cm]{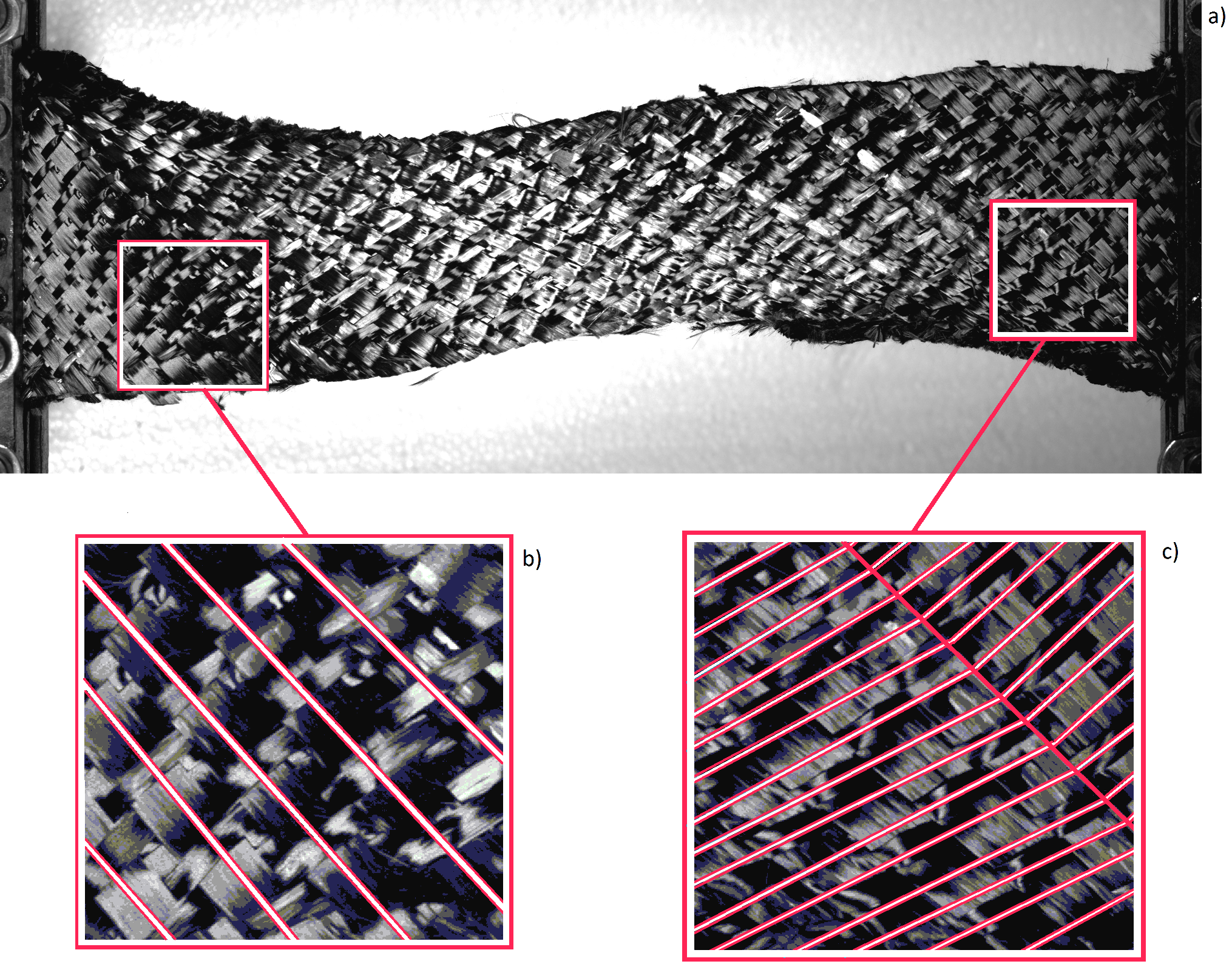}\protect\caption{\label{fig:Differential_Bending}Differential bending of the thick
(b) and thin (c) yarns as observed in the experimental deformed shape
(a).}
\end{figure}

We will more precisely describe later which is the precise pattern
of the yarns inside the considered specimen. On the other hand, we
want to point out here that, indeed, a sensible \textit{\textcolor{black}{differential
bending}} can be observed in the considered specimen. In fact, the
thin yarns are seen to be sensibly bent in some thin transition layers,
while the thick yarns do not bend at all as highlighted in Fig. \ref{fig:Differential_Bending}.
 Such differential bending is certainly at the origin of the asymmetric
deformed shape.

\subsection{\label{sub:Equil_2nd_Grad}The equilibrium problem for an orthotropic
micromorphic continuum subjected to Bias Extension Test}

In this subsection we want to set up a consistent continuum mechanical
framework which is able to account for 
  
\begin{itemize}
 
\item the presence of t\textit{\textcolor{black}{wo preferred material directions}}
inside the macroscopic body,
\item the unbalance of the fabric which, for what said in subsection \ref{sub:The-physical-problem:},
basically means that the two families of yarns have \textit{\textcolor{black}{different
bending stiffnesses}},
\item the \textit{\textcolor{black}{relative slipping}} of the yarns.  

\end{itemize}
 
In order to do so, we need to remark that the desired continuum model
(constrained micromorphic) must be such that
  
\begin{itemize}
 
\item the chosen first gradient constitutive laws account for the \textit{\textcolor{black}{orthotropy}}
of the considered medium,
\item it allows to describe the \textit{\textcolor{black}{curvature}} of
material lines. Moreover, the resistance to curvature must be different
for the two family of yarns, which means that the introduced constitutive
laws must account for different bending stiffnesses for the two families
of yarns. In order to reduce the adopted micromorphic model to a second
gradient one, suitable constrains must be introduced in order to let
higher order derivatives appear in the strain energy density,
\item notwithstanding the quasi-inextensibility of the carbon yarns, it
includes equivalent elongation modes in the yarns directions in order
to indirectly account for the slipping of the fibers. The fact of
considering such elongation modes, also allows the model the possibility
of describing the initial decrimping of the yarns.  

\end{itemize}
 
\medskip{}

In order to formulate the equilibrium problem for the considered continuum
subjected to a Bias Extension Test in the general framework introduced
in section \ref{sec:The-Principle-of}, we need to specify the particular
form that $\mathcal{P}^{int}$, $\mathcal{P}^{ext}$, $Q$ , $D$,
$T_{u}$ and $T_{\varphi}$ take in the treated example.

\subsubsection{\label{sub:Power-of-internal_Lag_mult}Work of internal actions for
a 2D, orthotropic, constrained micromorphic continuum}

We start by specifying the expression of the work of internal actions
which is suitable to describe the deformation of the considered system.
To do so, we start by recalling classical results for first gradient,
hyperelastic orthotropic continua which prescribe the functional dependence
that the strain energy density of an orthotropic continuum must have
on the Cauchy-Green strain tensor $C=F^{T}\cdot F$ (see e.g. \cite{Boheler,Boheler1,Phil1,Holzapfel book,Itskov,Itskov 1,Ogden,Ogden CISM,Raoult Annie,Spencer,David6}). 

Here and in the sequel, we denote by $\chi:B\rightarrow\mathbb{R}^{3}$
the placement function associated to the considered body that associates
to any material particle $X\in B$ its current position $x$ in the
deformed configuration and which can be related to the displacement
field by means of the relation $\chi=u+X$. Moreover, we denote by
$F=\nabla\chi$ the space gradient of the introduced placement field.

\medskip{}

We start by assuming that the strain energy density can be given in
the form
\begin{equation}
W(C,\nabla\varphi)=W_{I}(C)+W_{II}(\nabla\varphi).\label{eq:W_additive}
\end{equation}

Representation theorems for 3D orthotropic first gradient materials
are available in the literature (see e.g. \cite{Raoult Annie,Itskov0,Cuomo}),
which state that the functional dependence of the strain energy density
on the strain tensor $C$ must be given in terms of its invariants
$i_{O}:=\left\{ i_{1},i_{4},i_{6},i_{8},i_{9},i_{10}\right\} $, where
the introduced invariants are defined in table \ref{table: Invariants}
in which also their physical interpretation can be found. 

Explicit expressions for the strain energy potential as function of
the invariants $i_{O}$ which are suitable to describe the real behavior
of orthotropic hyperelastic materials are difficult to be found in
the literature. Certain constitutive models are for instance presented
in \cite{Itskov}, where some polyconvex energies for orthotropic
materials are proposed to describe the deformation of rubbers in uniaxial
tests. Explicit anisotropic hyperelastic potentials for soft biological
tissues are also proposed in \cite{Holzapfel} and reconsidered in
\cite{Neff1,NeffBis} in which their polyconvex approximations are
derived. Other examples of polyconvex energies for anisotropic solids
are given in \cite{David5}. It is even more difficult to find in
the literature reliable constitutive models for the description of
the real behavior of fibrous composite reinforcements at finite strains
but some attempts can be for instance recovered in \cite{Aimene,Phil1}.
Furthermore, the mechanical behavior of composite preforms with rigid
organic matrix (see e.g. \cite{Yves1,Yves3,Yves4,Yves2}) is quite
different from the behavior of the sole fibrous reinforcements (see
e.g. \cite{Phil3D}) rendering the mechanical characterization of
such materials a major scientific and technological issue. 

\begin{table}[h]
\begin{centering}
\begin{tabular}{|c|c|c|}
\hline 
Invariant & Expression & Meaning in terms of deformation\tabularnewline
\hline 
\hline 
$i_{1}$ & $\mathrm{tr}(C)$ & Averaged changes of length\tabularnewline
\hline 
$i_{4}$ & $m_{1}\cdot C\cdot m_{1}$ & Local stretch in the direction $\ensuremath{m_{1}}$ \tabularnewline
\hline 
$i_{6}$ & $m_{2}\cdot C\cdot m_{2}$ & Local stretch in the direction $\ensuremath{m_{2}}$ \tabularnewline
\hline 
$i_{8}$ & $m_{1}\cdot C\cdot m_{2}$ & Angle variation between the directions $\ensuremath{(m_{1},m_{2})}$\tabularnewline
\hline 
$i_{9}$ & $m_{1}\cdot C\cdot m_{3}$ & Angle variation between the directions $\ensuremath{(m_{1},m_{3})}$ \tabularnewline
\hline 
$i_{10}$ & $m_{2}\cdot C\cdot m_{3}$ & Angle variation between the directions $\ensuremath{(m_{2},m_{3})}$\tabularnewline
\hline 
\end{tabular}
\par\end{centering}

\protect\caption{Invariants of the Green-Lagrange strain tensor in the orthotropic
case. The vectors $m_{1}$ and $m_{2}$ are unit vectors in the two
privileged directions of the material and $m_{3}:=m_{1}\times m_{2}$.
\label{table: Invariants}}
\end{table}

\medskip{}
For the particular 2D case that we study here we assume that the first
gradient energy takes the following particular constitutive form
\begin{equation}
W_{I}(C)=\frac{1}{2}K_{el}\left[(\sqrt{i_{4}}-1)^{2}+(\sqrt{i_{6}}-1)^{2}\right]+\frac{1}{2}K_{sh}i_{8}^{2},\label{eq:W_I}
\end{equation}
where the parameters $K_{el}$ and $K_{sh}$ are assumed to be constant.
The constitutive choice (\ref{eq:W_I}) can be motivated by noticing
that the invariants $i_{9}$ and $i_{10}$ represent out-of-plain
angle variations of the yarns which are not considered in the present
2D case. Moreover, in a first instance, the isotropic invariant $i_{1}$
can be considered to be uninfluential in the considered orthotropic
case. 

It must be remarked that this simple quadratic choice for the first
gradient strain energy density, even if providing geometric non-linearities,
could be not sufficiently general to describe larger deformations
for which more complex hyperelastic constitutive laws should be introduced.
More than that, since for very large strains the integrity of the
material starts to be affected due to the excessive slipping, there
is no interest in attempting the modeling of the targeted unbalanced
materials after a given strain threshold. In order to precisely identify
the maximum strain that the material can withstand before failure
due to excessive slipping, more experimental campaigns should be carried
out. We limit ourselves here to remark that: 
  
\begin{itemize}
 
\item After a first threshold the material behavior starts to present a
softening (see experimental pattern in Fig. \ref{fig:SIM-LD}) which
can be directly related to slipping. To model such behavior, more
general hyperelastic laws with respect to the one presented in this
paper should be introduced, but this falls outside the scope of the
present work. 
\item If the experiment is prolonged, the slipping becomes so important
that the integrity of the material starts to be affected and some
yarns are pulled out of the specimen. In this case, both discrete
and continuum model loose their predictability.  

\end{itemize}
 
In summary, we are saying that, with the considered constitutive choice
and when remaining in the moderate strain regime, the main first gradient
deformation modes allowed in the deformation of the considered material
are 
  
\begin{itemize}
 
\item the angle variation $i_{8}$ between the warp and weft direction
\item the equivalent elongations $i_{4}$ and $i_{6}$ in the directions
of the warp and weft which account for decrimping and, eventually
for slipping.  

\end{itemize}
 
\medskip{}

As for the micromorphic energy, we make the following particular constitutive
choice
\begin{equation}
W_{II}(\nabla\varphi)=\frac{1}{2}\left\langle \alpha,\nabla\varphi\right\rangle ^{2}=\frac{1}{2}\left(\alpha_{1}\varphi_{,1}+\alpha_{2}\varphi_{,2}\right),\label{eq:W_II}
\end{equation}
where the vector $\alpha=(\alpha_{1},\alpha_{2})$ is the vector of
constant micromorphic elastic parameters whose components $\alpha_{1}$
and $\alpha_{2}$ have to be different to account for the unbalance
of the microscopic characteristics of the material. Moreover, we denoted
by $\varphi_{,1}$ and $\varphi_{,2}$ the space derivatives of $\varphi$
with respect to the space coordinates in the directions $m_{1}$ and
$m_{2}$ (warp and weft) respectively.

\medskip{}

We explicitly remark at this point that we want to give a precise
physical meaning to the micro-descriptor $\varphi$ in order to catch
at best the experimental behaviors described in subsection \ref{sub:The-physical-problem:}.
We have pointed out that a supplementary deformation mode with respect
to classical first gradient ones must be introduced in order to catch
all the physics of the considered problem. More particularly, additionally
to angle variations ($i_{8}$) and equivalent elongations in the yarns'
directions ($i_{4}$ and $i_{6}$), we need to account for the bending
of the two families of fibers. It is known (see also \cite{dell_Steigmann,MadeoBias})
that bending strains of the fibers inside the considered macroscopic
specimen can be accounted for by introducing second derivatives of
the displacement field in the strain energy density. In particular,
such bending of the yarns can be related to the space gradient of
the angle variation $i_{8}$: if sharp variations of angle between
warp and weft occur in small regions inside the specimen, it means
that the yarns must necessarily bend in order to rapidly change their
direction and give rise to such angle variation. 

In the light of such remarks, it is sensible to suppose that the introduced
micro-descriptor $\varphi$ must be indeed related to the angle variation
$i_{8}$: if, for example, we let $\varphi$ tend to $i_{8}$, then
expression (\ref{eq:W_II}) for the strain energy density accounts
for space derivatives of the angle variation and hence, finally, for
the bending of the two families of yarns. Indeed, (see also \cite{dell_Steigmann,MadeoBias}),
it may be rather easily inferred how $i_{8,1}$ can be interpreted
as the bending of the yarns disposed in the $m_{1}$ direction and,
analogously, $i_{8,2}$ represents the bending of the yarns aligned
in the $m_{2}$ direction.

Based on the physics of the problem discussed in the previous subsection,
we do not introduce second gradient effects related to the gradients
of the other invariants. We are then excluding that sharp spacial
changes of elongation occur in the considered material.

\medskip{}

At this point, the reader may believe that the fact of considering
a micromorphic medium is redundant for treating the considered problem
of the Bias Extension Test, since a second gradient model could have
been directly introduced, instead of constraining a micromorphic model
to become a second gradient one. Nevertheless, the intermediary step
of passing through a micromorphic model is essential, at least for
two reasons
  
\begin{itemize}
 
\item the imposed boundary conditions take a precise and unique meaning
\item the numerical implementation of the considered problem is more easily
treatable since lower order differential equations are involved.  

\end{itemize}
 
The first point of the unique meaning of the imposed boundary conditions
is crucial if one wants to deal with a model which has an easily recognizable
physically grounded interpretation. In fact, as far as second gradient
theories are concerned, the boundary conditions that can be imposed
may take different, but equally legitimate, forms for the same physical
problem (see e.g. \cite{incomplete}): for example, in a second gradient
theory, a given angle can be imposed either by directly assigning
the angle or by suitably choosing the components of the normal derivative
of displacement on the boundary. Depending on whether one choice of
the kinematical conditions or the other one is made, the dual traction
counterparts (dual of the angle variation or of the normal derivative)
have different expressions and the definition of the force can be
also shown to be non-equivalent in the two cases. Such non-uniqueness
of the way of imposing second gradient boundary conditions is directly
related to the number of integration by parts which one decides to
make in the expression of the internal work: in second gradient theories,
the second gradient of the virtual displacement can be integrated
by parts twice, by making use of the standard divergence theorem and
of the surface divergence theorem.

On the other hand, micromorphic models only involve first gradients
of the introduced kinematical fields, so that only one level of integration
by parts can be conceived (only the standard divergence theorem is
used in a micromorphic model). This fact, avoids any sort of indeterminacy
for the imposable boundary conditions when micromorphic models are
considered (see e.g. also \cite{Bleustein}).

In order to implement the fact that we want to constrain the micromorphic
energy (\ref{eq:W_additive}) to a second gradient one, we need to
introduce a suitable Lagrange multiplier $\Lambda$ with associated
strain energy density $W_{\Lambda}$ 
\begin{equation}
W_{\Lambda}(\Lambda,i_{8},\varphi)=\Lambda\left(\varphi-i_{8}\right)\label{eq:W_II-1}
\end{equation}

In this constrained framework the global energy of the system is not
simply (\ref{eq:W_additive}), but must be complemented with this
additional coupling term and in the considered 2D case becomes
\begin{equation}
W(i_{4},i_{6},i_{8},\varphi,\nabla\varphi,\Lambda)=W_{I}(i_{4},i_{6},i_{8})+W_{II}(\nabla\varphi)+W_{\Lambda}(\Lambda,i_{8},\varphi).\label{eq:W_additive-1}
\end{equation}
\textcolor{black}{We explicitly mention that, the introduction of
the Lagrange multiplier $\Lambda$ does not modify the definition
of the set $D$.}

\medskip{}

According to the constitutive choices (\ref{eq:W_I}) and (\ref{eq:W_II-1}),
the work of internal actions of the considered 2D, orthotropic, constrained
micromorphic continuum can be written as
\[
\mathcal{P}_{II}^{int}=\delta\mathcal{A}^{int}=-\delta\int_{B}\left(W_{I}+W_{II}+W_{\Lambda}\right)\,dv=-\int_{B}\left(\delta W_{I}+\delta W_{II}+\delta W_{\Lambda}\right)\,dv.
\]
Computing the first variation of the introduced internal action functional,
we can write
\begin{align*}
\mathcal{P}_{II}^{int} & =-\int_{B}\left[\frac{\partial W_{I}}{\partial i_{4}}\delta i_{4}+\frac{\partial W_{I}}{\partial i_{6}}\delta i_{6}+\left(\frac{\partial W_{I}}{\partial i_{8}}+\frac{\partial W_{\Lambda}}{\partial i_{8}}\right)\delta i_{8}+\frac{\partial W_{II}}{\partial\nabla\varphi}\delta\nabla\varphi+\frac{\partial W_{\Lambda}}{\partial\varphi}\delta\varphi+\frac{\partial W_{\Lambda}}{\partial\Lambda}\delta\Lambda\right]\,dv,\\
\end{align*}
which for the particular constitutive choice made to treat the physical
example considered in this paper, takes the particular form

\begin{align}
\mathcal{P}_{II}^{int} & =\int_{B}\left[-\frac{1}{2}K_{el}\left(1-\frac{1}{\sqrt{i_{4}}}\right)\delta i_{4}-\frac{1}{2}K_{el}\left(1-\frac{1}{\sqrt{i_{6}}}\right)\delta i_{6}-\left(K_{sh}i_{8}-\Lambda\right)\delta i_{8}\right.\nonumber \\
\label{eq:P2_int}\\
 & \left.-{\color{white}\frac{1}{2}}\left\langle \alpha,\nabla\varphi\right\rangle \left\langle \alpha,\nabla\delta\varphi\right\rangle -\Lambda\,\delta\varphi-(\varphi-i_{8})\delta\Lambda\right]\,dv.\nonumber 
\end{align}
Considering the definitions of the invariants given in table (\ref{table: Invariants}),
it can be checked that
\begin{align*}
\delta i_{4} & =m_{1}\cdot\delta\left(F^{T}\cdot F\right)\cdot m_{1}=2\left(F\cdot m_{1}\right)\cdot\left(\delta F\cdot m_{1}\right)\\
\\
\delta i_{6} & =m_{2}\cdot\delta\left(F^{T}\cdot F\right)\cdot m_{2}=2\left(F\cdot m_{2}\right)\cdot\left(\delta F\cdot m_{2}\right)\\
\\
\delta i_{8} & =m_{1}\cdot\delta\left(F^{T}\cdot F\right)\cdot m_{2}=\left(F\cdot m_{2}\right)\cdot\left(\delta F\cdot m_{1}\right)+\left(F\cdot m_{1}\right)\cdot\left(\delta F\cdot m_{2}\right),
\end{align*}
so that, replacing such expressions in the first three terms of the
internal work (\ref{eq:P2_int}), recalling that $\delta F=\nabla(\delta\chi)=\nabla(\delta u)$,
suitably integrating by parts each term and using the divergence theorem
we finally get

\begin{alignat}{1}
\mathcal{P}_{II}^{int} & =\int_{B}\ \left\langle \mathrm{Div}\left[K_{el}\:\left(1-\frac{1}{\sqrt{i_{4}}}\right)\left(F\cdot m_{1}\otimes m_{1}\right)\right]+\mathrm{Div}\left[K_{el}\:\left(1-\frac{1}{\sqrt{i_{6}}}\right)\left(F\cdot m_{2}\otimes m_{2}\right)\right]\ ,\ \delta u\right\rangle \ dv\nonumber \\
\nonumber \\
 & +\int_{B}\ \left\langle \ \mathrm{Div}\left[\ \left(K_{sh}\:i_{8}-\Lambda\right)\:F\cdot\left(m_{1}\otimes m_{2}+m_{2}\otimes m_{1}\right)\ \right],\ \delta u\ \right\rangle \ dv-\int_{B}(\varphi-i_{8})\delta\Lambda\:dv\nonumber \\
\nonumber \\
 & -\int_{\partial B}\ \left[K_{el}\:\left(1-\frac{1}{\sqrt{i_{4}}}\right)\left(m_{1}\cdot n\right)+\left(K_{sh}\:i_{8}-\Lambda\right)\left(m_{2}\cdot n\right)\right]\ \left\langle \ \left(F\cdot m_{1}\right)\ ,\ \delta u\ \right\rangle \ ds\label{eq:P2_int-1}\\
\nonumber \\
 & -\int_{\partial B}\ \left[K_{el}\:\left(1-\frac{1}{\sqrt{i_{6}}}\right)\left(m_{2}\cdot n\right)+\left(K_{sh}\:i_{8}-\Lambda\right)\left(m_{1}\cdot n\right)\right]\ \left\langle \ \left(F\cdot m_{2}\right)\ ,\ \delta u\ \right\rangle \ ds\nonumber \\
\nonumber \\
 & +\int_{B}\ \left[\alpha\cdot\nabla\left(\,\left\langle \ \alpha,\nabla\varphi\ \right\rangle \,\right)-\Lambda\right]\ \delta\varphi-\int_{\partial B}\left\langle \ \alpha,\nabla\varphi\ \right\rangle \left\langle \ \alpha,n\ \right\rangle \ \delta\varphi\nonumber 
\end{alignat}
where we denoted by $n$ the unit normal to the Lagrangian boundary
$\partial B$. We explicitly remark that imposing arbitrary variations
$\delta\Lambda$ implies the desired constraint $\varphi=i_{8}$.

The expression (\ref{eq:P2_int-1}) is the \textit{irreducible form
of the work of internal actions} for the considered constrained micromorphic
medium. We explicitly remark that, the boundary contact actions intervening
in the considered problem expend work on $\delta u$ (forces) and
on $\delta\varphi$ (double-forces). As far as the boundary $\partial B$
is concerned, it can be inferred from Eq. (\ref{eq:P2_int-1}) that
the dual quantity to the virtual displacement $\delta u$ contains
informations about the first gradient deformations $i_{4}$, $i_{6}$,
$i_{8}$ as well as the Lagrange multiplier $\Lambda$. We are in
some way saying that the internal force which would eventually balance
an externally applied force involve deformation mechanisms related
to elongations of the warp and weft, angle variations, but also (through
the Lagrange multiplier $\Lambda$) some microstructure-related deformation
modes. On the other hand, the internal double-force (couple) which
would eventually balance an externally applied double force, involve
deformation mechanisms associated to the local bending of the yarns
(first derivatives of the angle variation $\varphi=i_{8}$). 

Analogously to what done in subsection \ref{sub:Reaction_Force},
we could use this irreducible expression of the work of internal forces
to suitably calculate the reaction forces and double-forces on the
part of the boundary where we assign the kinematical constraints.
Nevertheless, since in our numerical simulations we only use the second
method to calculate such reactions (passing through the evaluation
of volume integrals), we will limit ourselves to present it in subsection
\ref{sub:Evaluation-of-reaction}.

\subsubsection{Work of external actions for the considered constrained micromorphic
continuum}

According to the procedure shown in the previous section, in order
to formulate the equilibrium problem for a given continuum, the work
of external actions must be given on the portion of the boundary $\Sigma_{T}=\Sigma_{T_{1}}\bigcup\Sigma_{T_{2}}$
where tractions are assigned. According to the irreducible expression
(\ref{eq:P2_int-1}) which we consider for the work of internal actions,
the external work that balances the boundary terms of the internal
one, must contain ``\textit{forces}'' $f$ that expend work on the
virtual displacement $\delta u$ and ``\textit{double-forces}''
$\tau$ that expend work on the virtual angle variation $\delta\varphi$.
In particular, the work of external forces for the considered micromorphic
continuum is assumed to take the form
\begin{equation}
\mathcal{P}_{II}^{ext}=\int_{\Sigma_{T_{1}}}\left\langle f,\delta u\right\rangle \,ds+\int_{\Sigma_{T_{2}}}\tau\:\delta\varphi\,ds.\label{eq:P2_ext}
\end{equation}
For the particular case of the bias extension test considered here,
we have that traction boundary conditions are assigned on the surfaces
$\Sigma_{3}$ and $\Sigma_{4}$. More precisely, we have that $\Sigma_{T_{1}}\equiv\Sigma_{T_{2}}=\Sigma_{3}\bigcup\Sigma_{4}$:
forces and double forces are simultaneously assigned on $\Sigma_{3}$
and $\Sigma_{4}$. More particularly, since in the Bias Extension
Test $\Sigma_{3}$ and $\Sigma_{4}$ are free boundaries the assigned
value of forces and double forces is the null value:
\[
f=0,\qquad\tau=0.
\]

\subsubsection{Space of configurations for the bias extension test}

Once that the works of external and internal actions have been given
for the considered particular case, the kinematical boundary conditions
must be assigned and the associated space of configurations and of
admissible variations must be identified.

For the experimental set-up of the Bias Extension test shown in Fig.
\ref{fig:Experimental_Set_up}, we want to assign
  
\begin{itemize}
 
\item $u=0$ and $\varphi=0$ on $\Sigma_{1}$ and 
\item $u=u_{0}=\mathrm{const}$ and $\varphi=0$ on $\Sigma_{2}$. If we
want to frame this situation in the more general picture given in
section \ref{sec:The-Principle-of}, we have to set $\Sigma_{K_{1}}\equiv\Sigma_{K_{2}}=\Sigma_{1}\bigcup\Sigma_{2}$
and the space of configurations $Q\times D$ for the considered micromorphic
continuum subjected to a Bias Extension Test takes the particular
form
\begin{align}
Q & =\left\{ u\ \mid\ u=0\ \ \text{on\ }\ \Sigma_{1}\quad\text{and }\quad u=u_{0}=\mathrm{const},\ \ \text{on}\ \ \Sigma_{2}\right\} ,\nonumber \\
\label{eq:SC_2nd}\\
D & =\left\{ \varphi\ \mid\ \varphi=0\ \ \text{on\ }\ \Sigma_{1}\quad\text{and }\quad\varphi=0,\ \ \text{on}\ \ \Sigma_{2}\right\} \nonumber 
\end{align}

\end{itemize}
 
The spaces of admissible variations that must be associated to such
space of configurations are given by
\begin{equation}
T_{u}=\left\{ \delta u\ \mid\:u+\delta u\in Q\right\} ,\qquad T_{\varphi}=\left\{ \delta\varphi\ \mid\:\varphi+\delta\varphi\in D\right\} .\label{eq:VA_2nd}
\end{equation}
Since both the displacement $u$ and the angle variation $\varphi$
are assigned on $\Sigma_{1}\bigcup\Sigma_{2}$, then it must be $\delta u=0$
and $\delta\varphi=0$, so that the spaces of admissible variations
can equivalently be written as $T_{u}=\left\{ \delta u\ \mid\delta u=0\quad\text{on}\ \Sigma_{1}\bigcup\Sigma_{2}\right\} $
and $T_{\varphi}=\left\{ \delta\varphi\ \mid\delta\varphi=0\quad\text{on}\ \Sigma_{1}\bigcup\Sigma_{2}\right\} $.

Suitably generalizing what done before, the equilibrium problem for
the considered constrained micromorphic continuum subjected to a Bias
Extension Test can be formulated as

\textit{\textcolor{black}{Find $\left(u^{*},\varphi^{*}\right)\in Q\times D$
such that $\mathcal{P}^{int}(u^{*},\varphi^{*},\delta u,\delta\varphi)+\mathcal{P}^{ext}(u^{*},\varphi^{*},\delta u,\delta\varphi)=0\quad$
for any $\left(\delta u,\delta\varphi\right)\in T_{u^{*}}\times T_{\varphi^{*}}$,}}

where now $\mathcal{P}^{int}$, $\mathcal{P}^{ext}$, $Q$, $D$,
$T_{u}$ and $T_{\varphi}$ are given by (\ref{eq:P2_int-1}), (\ref{eq:P2_ext}),
(\ref{eq:SC_2nd}) and (\ref{eq:VA_2nd}) respectively.

The mathematical question of well-posedness of the geometrically nonlinear
micromorphic approach has been discussed in \cite{Neff1-Microm,Neff2-Microm,Neff3-Microm}-
the extension of these results to the anisotropic setting is straightforward.
Some attendant results in the large strain and small strain setting,
including an efficient numerical treatment and further modeling and
well-posedness results can be found in \cite{NeffMicromE,NeffMicromC,NeffMicromA,NeffMicromB,NeffMicromD,NeffMicromF}.

\subsubsection{\label{sub:Evaluation-of-reaction}Evaluation of reaction forces
and double forces}

As done in subsection \ref{sub:Reaction_Force} we can ask ourselves
how, in the framework of the considered constrained micromorphic model,
we can calculate the reaction forces and double-forces which would
be needed to produce the kinematical conditions assigned on $\Sigma_{1}$
and $\Sigma_{2}$ as well as the solution $\left(u^{*},\varphi^{*}\right)$
obtained solving the considered constrained micromorphic equilibrium
problem. To do so, we proceed as in subsection \ref{sub:Reaction_Force}
and we introduce a work of external forces also on the portion of
the boundary $\Sigma_{1}\bigcup\Sigma_{2}$ where kinematical boundary
conditions have been assigned, so that the work of external forces
takes the modified form
\begin{equation}
\mathcal{P}_{II}^{ext}=\int_{\Sigma_{T_{1}}}\left\langle f,\delta u\right\rangle \,ds+\int_{\Sigma_{T_{2}}}\tau\:\delta\varphi\,ds+\int_{\Sigma_{K_{1}}}\left\langle f_{R},\delta u\right\rangle ds+\int_{\Sigma_{K_{2}}}\tau_{R}\,\delta\varphi ds,\label{eq:Pext_2_mod}
\end{equation}
where we remind that $\Sigma_{T_{1}}\equiv\Sigma_{T_{2}}=\Sigma_{3}\bigcup\Sigma_{4}$
and that $f=0$ and $\tau=0$ for the considered Bias Extension Test.
Moreover, $\Sigma_{K_{1}}\equiv\Sigma_{K_{2}}=\Sigma_{1}\bigcup\Sigma_{2}$
. In order to determine the reaction forces and double forces which
balance the imposed displacements and angle variations on $\Sigma_{1}$
and $\Sigma_{2}$, we could pass through the use of the irreducible
form of the work of internal actions, as we have explicitly shown
for the first gradient case in subsection \ref{sub:Reaction_Force}. 

Nevertheless, since for the performed numerical simulations we only
used the evaluation of volume integrals, we present here only this
last method. To this aim, we write the Principle of Virtual Work for
the considered particular example, evaluated in the solution $\left(u^{*},\varphi^{*},\Lambda^{*}\right)$
of the equilibrium problem, as
\begin{gather}
\int_{B}\left[-\frac{1}{2}K_{el}\left(1-\frac{1}{\sqrt{i_{4}^{*}}}\right)\delta i_{4}-\frac{1}{2}K_{el}\left(1-\frac{1}{\sqrt{i_{6}^{*}}}\right)\delta i_{6}-\left(K_{sh}i_{8}^{*}-\Lambda^{*}\right)\delta i_{8}-\left\langle \alpha,\nabla\varphi^{*}\right\rangle \left\langle \alpha,\nabla\delta\varphi\right\rangle -\Lambda^{*}\,\delta\varphi\right]\,dv\nonumber \\
\label{eq:PPV_2nd_ustar}\\
-\int_{B}(\varphi^{*}-i_{8}^{*})\delta\Lambda\:dv+\int_{\Sigma_{K_{1}}}\left\langle f_{R},\delta u\right\rangle ds+\int_{\Sigma_{K_{2}}}\tau_{R}\,\delta\varphi=0\nonumber 
\end{gather}
where the expression for the work of internal actions given in Eq.
(\ref{eq:P2_int}) has been used setting $i_{4}^{*}:=i_{4}(u^{*})$,
$i_{6}^{*}:=i_{6}(u^{*})$, $i_{8}^{*}:=i_{8}(u^{*})$ and where we
impose that this last form of the Principle of Virtual Work must be
satisfied for any \textit{$\delta u$, $\delta\varphi$ }and\textit{
$\delta\Lambda$ .}

In order to calculate the reaction force, we choose particular test
functions $\delta\bar{u}$, $\delta\bar{\varphi}$ and $\delta\bar{\Lambda}$
such that
  
\begin{enumerate}
 
\item $\delta\bar{u}$ is vanishing on $\Sigma_{1}$ non-vanishing and constant
on $\Sigma_{2}$
\item $\delta\bar{u}$ is an arbitrarily assigned, non-vanishing continuous
function outside $\Sigma_{K_{1}}$
\item $\delta\bar{\varphi}=0$ everywhere
\item $\delta\bar{\Lambda}=0$ everywhere.  

\end{enumerate}
 
With this choice, we can evaluate the reaction force as

\begin{gather}
\left\langle R,\delta\bar{u}\right\rangle =\int_{B}\left[\frac{1}{2}K_{el}\left(1-\frac{1}{\sqrt{i_{4}^{*}}}\right)\delta\bar{i}_{4}+\frac{1}{2}K_{el}\left(1-\frac{1}{\sqrt{i_{6}^{*}}}\right)\delta\bar{i}_{6}+\left(K_{sh}i_{8}^{*}-\Lambda^{*}\right)\delta\bar{i}_{8}\right]\,dv\label{eq:reacMicrom}
\end{gather}
where we set $R:=\int_{\Sigma_{2}}f_{R}\,ds$, $\delta\bar{i}_{4}:=\delta i_{4}(\delta\bar{u})$,
$\delta\bar{i}_{6}:=\delta i_{6}(\delta\bar{u})$ and $\delta\bar{i}_{8}:=\delta i_{4}(\delta\bar{u})$.
In order to evaluate the three components of the reaction force, we
choose suitable test functions $\delta\bar{u}$ which are aligned
with one of the directions of the used reference system. In particular,
let $\left\{ e_{i}\right\} _{i\in\left\{ 1,2,3\right\} }$, be an
orthonormal basis with respect to which we want to evaluate the three
components $R_{i}=\left\langle R,e_{i}\right\rangle $ of the reaction
force $R$. We choose a test function $\delta\bar{u}$ which possesses
all the characteristics previously listed except that the point 1.
is replaced by 
  
\begin{itemize}
 
\item $\delta\bar{u}$ is equal to $e_{i}$ on $\Sigma_{1}$ .  

\end{itemize}
 
In this case the component $R_{i}$ can be easily calculated according
to Eq. (\ref{eq:reacMicrom}).

\medskip{}

Analogously, in order to calculate the reaction double-force, we can
choose particular test functions $\delta\bar{u}$, $\delta\bar{\varphi}$
and $\delta\bar{\Lambda}$ such that
  
\begin{enumerate}
 
\item $\delta\bar{u}=0$ everywhere
\item $\delta\bar{\varphi}$ is vanishing on $\Sigma_{1}$, non-vanishing
and constant on $\Sigma_{2}$
\item $\delta\bar{\varphi}$ is an arbitrarily assigned, non-vanishing continuous
function outside $\Sigma_{K_{1}}$
\item $\delta\bar{\Lambda}=0$ everywhere.  

\end{enumerate}
 
With this choice, we can evaluate the reaction double-force as
\begin{gather}
\mathcal{T}\,\delta\bar{\varphi}=\int_{B}\left[-\left\langle \alpha,\nabla\varphi^{*}\right\rangle \left\langle \alpha,\nabla\delta\bar{\varphi}\right\rangle -\Lambda^{*}\,\delta\bar{\varphi}\right]\,dv,\label{eq:ReacMicrom2}
\end{gather}
where we set $\mathcal{T}:=\int_{\Sigma_{K_{2}}}\tau_{R}\,ds$.

\subsection{Numerical simulations for the constrained continuum micromorphic
model}

In this subsection we show the numerical simulations that we have
performed in order to find numerical solutions of the equilibrium
problem for an unbalanced fabric formulated as the equilibrium of
a constrained micromorphic continuum subjected to a Bias Extension
Test as presented in subsection \ref{sub:Equil_2nd_Grad}. Such equilibrium
problem has been implemented in COMSOL$^{\circledR}$ using the weak
form package in which the expression (\ref{eq:P2_int}) for the work
of internal actions can be explicitly given. The work of external
actions (\ref{eq:P2_ext}) is given in COMSOL$^{\circledR}$ just
leaving $\Sigma_{3}$ and $\Sigma_{4}$ free which means that forces
and double forces are assigned to be vanishing on the considered subsets
of the boundary. Finally, the space of configurations (\ref{eq:SC_2nd})
is given assigning the kinematical boundary conditions on $\Sigma_{1}$
and $\Sigma_{2}$ as explicitly established in subsection (\ref{sub:The-physical-problem:}).

The numerical values used for the constitutive parameters are shown
in table (\ref{table: Par}).

\begin{table}[h]
\begin{centering}
\begin{tabular}{|c|c|c|c|}
\hline 
$K_{el}$ & $K_{sh}$ & $\alpha_{1}$ & $\alpha_{2}$\tabularnewline
\hline 
\hline 
0.7 MPa & 21 kPa & 2 kPa$\cdot$m$^{2}$ & 0.02 kPa$\cdot$m$^{2}$\tabularnewline
\hline 
\end{tabular}
\par\end{centering}

\protect\caption{\label{table: Par}Parameters of the constrained micromorphic continuum
model.}
\end{table}
 Such values have been chosen following a precise heuristic procedure
that we present here
  
\begin{figure}[H]
\begin{centering}
\includegraphics[scale=0.7]{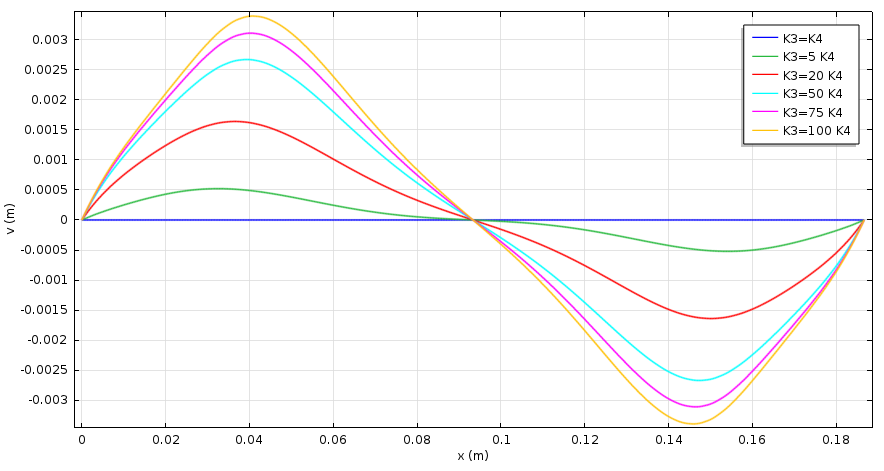}
\par\end{centering}

\protect\caption{\label{fig:SIM-alpha1}Vertical displacement of the mean horizontal
axis for an imposed displacement of 56 mm and different values of
$\alpha_{1}$. We remark that no distortion of the specimen is present
for a balanced fabric ($\alpha_{1}=\alpha_{2}$).}
\end{figure}

\begin{figure}[H]
\begin{centering}
\includegraphics[scale=0.5]{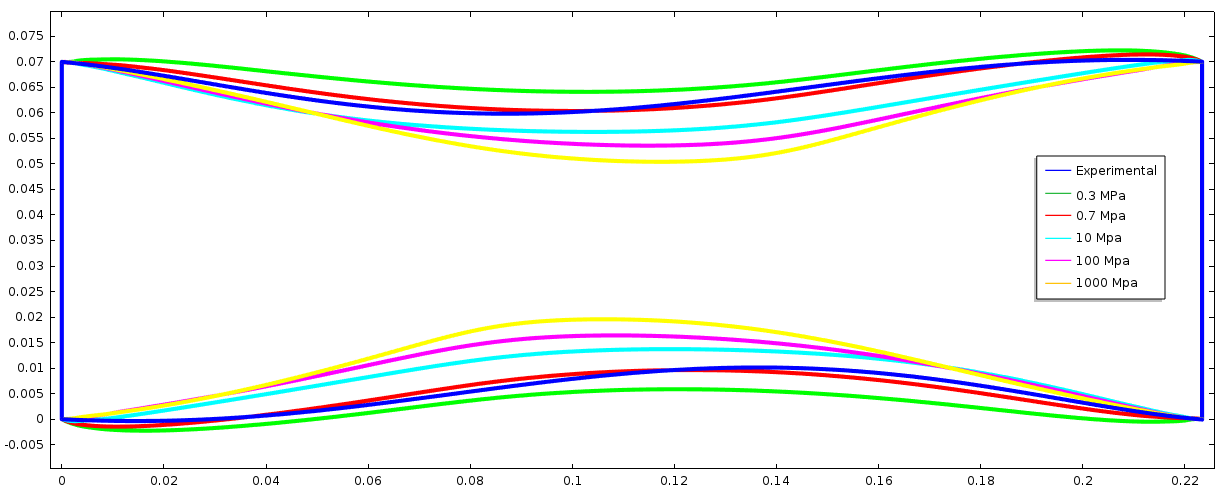}
\par\end{centering}

\protect\caption{\label{fig:SIM-Kel}Deformed shape for a displacement of 37 mm and
different values of $K_{el}$}
\end{figure}

\begin{figure}[H]
\begin{centering}
\includegraphics[scale=0.5]{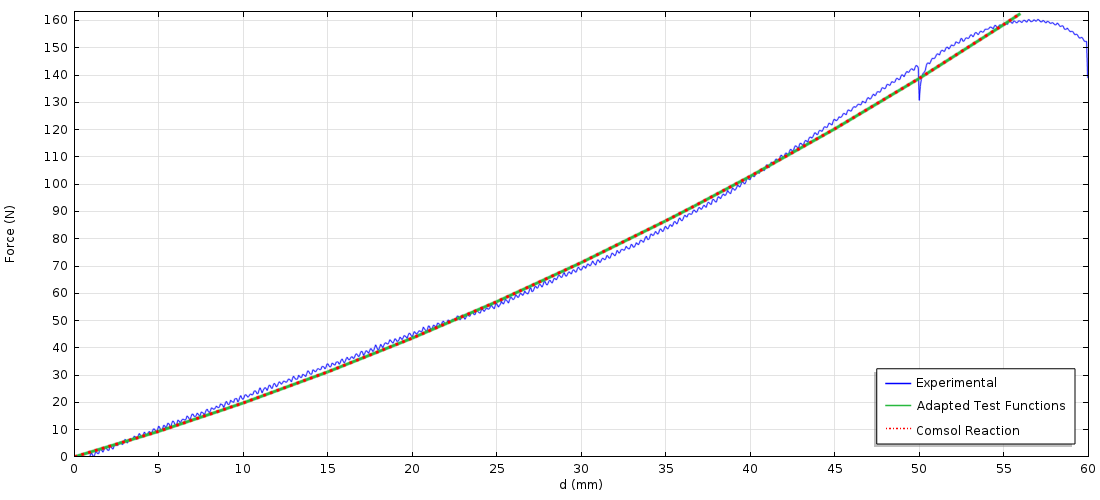}
\par\end{centering}

\protect\caption{\label{fig:SIM-LD} Load/displacement curve for the constrained micromorphic
model }
\end{figure}

\begin{itemize}
 
\item First of all we remarked the bending stiffness $\alpha_{2}$ of the
thin yarns is very small (eventually almost vanishing), so that we
choose a tentative (very small) value for such parameter
\item subsequently, we choose the bending stiffness $\alpha_{1}$ of the
thick yarns in order to fit at best the experimental S-shape of the
specimen (in Fig. \ref{fig:SIM-alpha1} it is shown as the unbalance
on the bending stiffnesses of warp and weft is responsible for the
S-shape of the specimen)
\item then, we tuned the value of the ``sliding'' parameter $K_{el}$
which was seen to have a direct influence on the in-plane thickness
of the specimen. Turning-on this parameter the height in the middle
of the specimen itself starts increasing and becomes closer to the
real experimental shape (see e.g. Fig. (\ref{fig:SIM-Kel}))
\item Finally, the in-plane shear parameter $K_{sh}$ is tuned in order
to fit at best the experimental load-displacement curve (see Fig.
\ref{fig:SIM-LD}).  

\end{itemize}
 
We need to spend some extra words on the way in which the reaction
force and double force have been numerically calculated. 
\begin{figure}[h]
\begin{centering}
\includegraphics[scale=0.9]{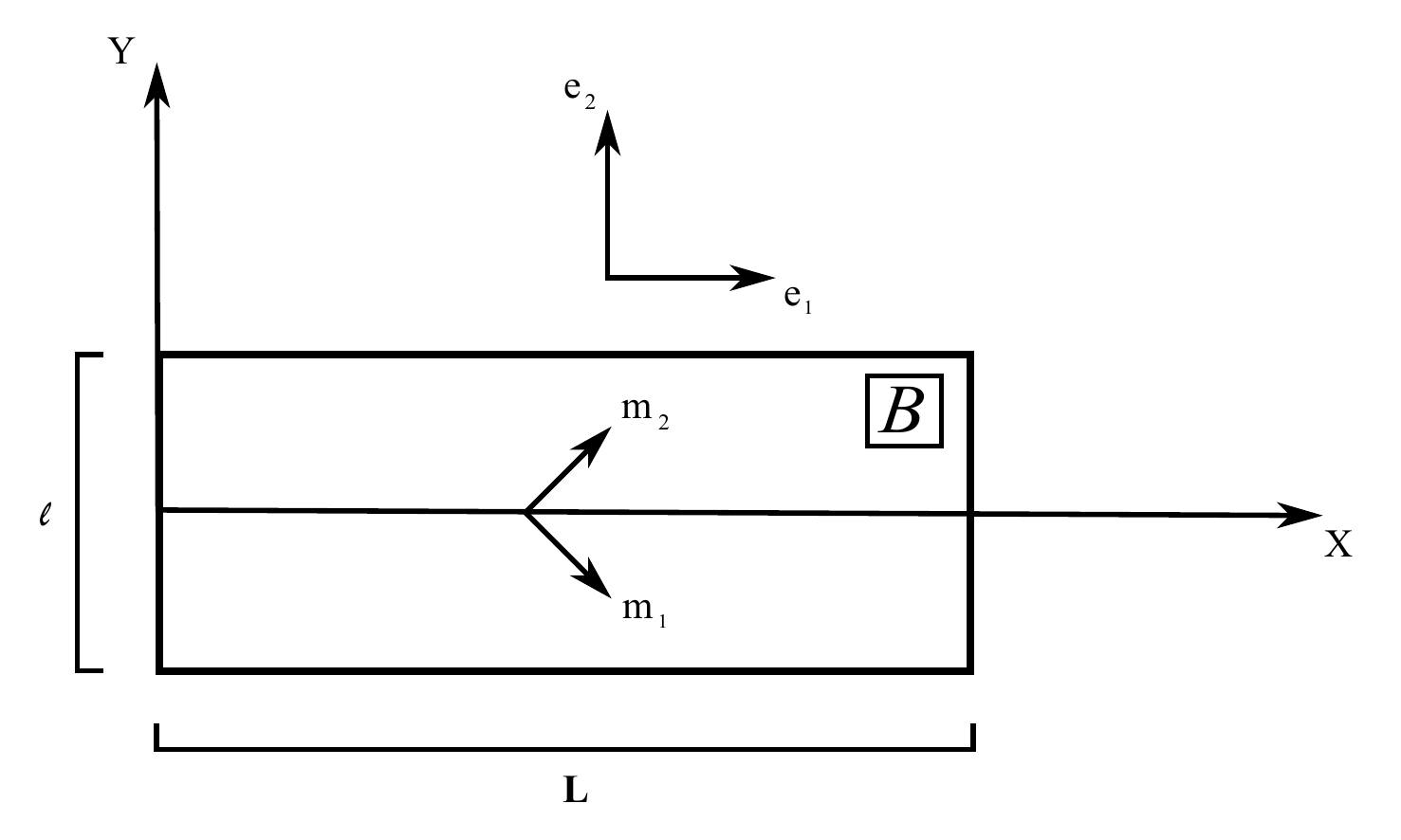}
\par\end{centering}

\protect\caption{\label{fig:Ref_Frame}Definition of the global ($\left\{ e_{1},e_{2}\right\} $)
and material ($\left\{ m_{1},m_{2}\right\} $) reference frames for
the BET.}

\end{figure}
We followed the procedure presented in subsection \ref{sub:Evaluation-of-reaction}
and, once calculated the solution $\left(u^{*},\varphi^{*},\Lambda^{*}\right)$
for each imposed displacement $u_{0}$, we calculated the horizontal
component of the reaction force $R$ with respect to an orthonormal
reference frame $\left\{ e_{1},e_{2}\right\} $ chosen as in Fig.
\ref{fig:Ref_Frame} by using equation (\ref{eq:reacMicrom}), where
we chose particular test functions $\delta\bar{u}$, $\delta\bar{\varphi}$
and $\delta\bar{\Lambda}$ such that
  
\begin{enumerate}
 
\item $\ \delta\bar{u}\cdot e_{1}=\frac{X}{L},\qquad\forall(X,Y)\in B$ 
\item $\delta\bar{u}\cdot e_{2}=0,\qquad\forall(X,Y)\in B$ 
\item $\delta\bar{\varphi}=0,\qquad\forall(X,Y)\in B$ 
\item $\delta\bar{\Lambda}=0,\qquad\forall(X,Y)\in B$.  

\end{enumerate}
 
The obtained force-displacement curve is shown in Fig. \ref{fig:SIM-LD},
where the comparison with the experimental curve and with the reaction
force automatically calculated by COMSOL$^{\circledR}$ are also depicted.

\begin{figure}[H]
\begin{centering}
\includegraphics[scale=0.5]{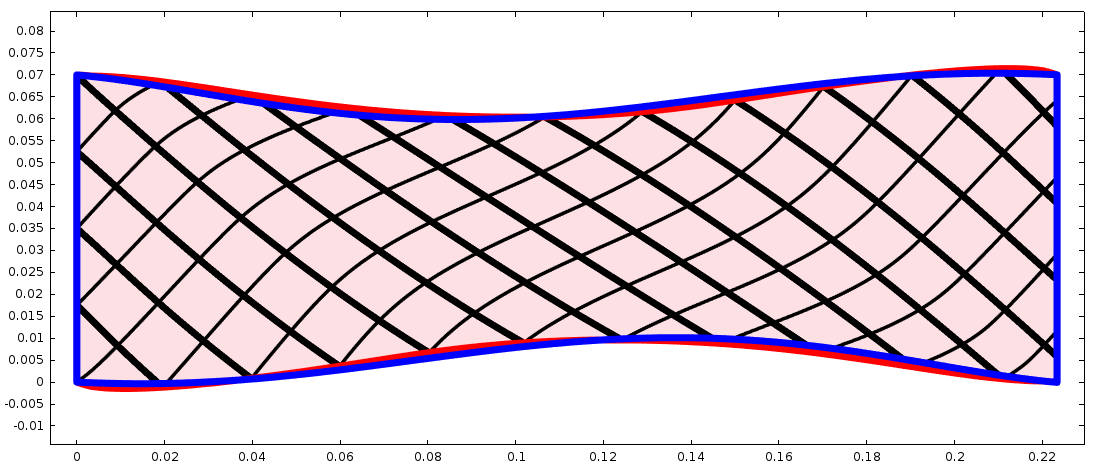}
\par\end{centering}

\protect\caption{Deformed shape in the simulation (red with black fibers) and experimental
(Blue) for a displacement of 37 mm \label{fig:SIM-Shape}}
\end{figure}

\begin{figure}[H]
\begin{centering}
\includegraphics[scale=0.5]{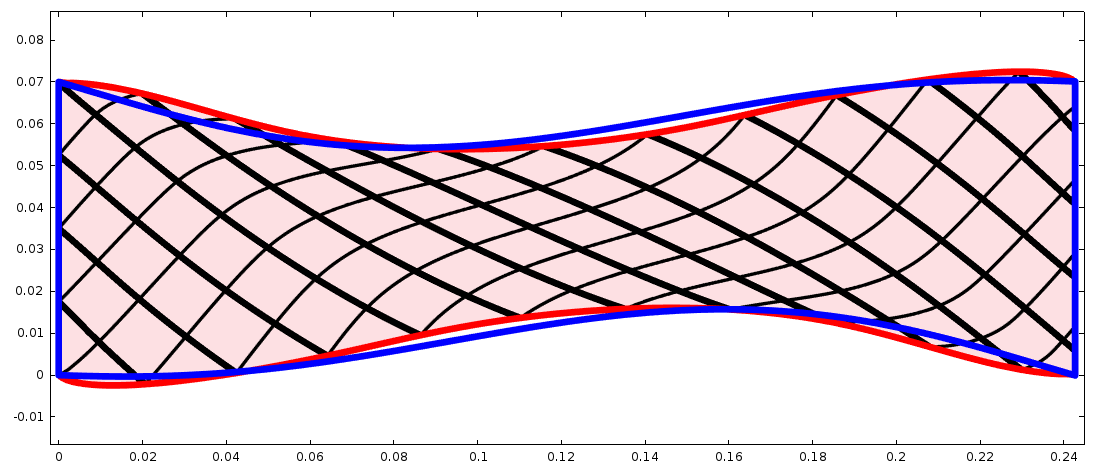}
\par\end{centering}

\protect\caption{Deformed shape in the simulation (red with black fibers) and experimental
(blue) for a displacement of 56 mm \label{fig:SIM-Shape-1}}
\end{figure}

In figures \ref{fig:SIM-Shape} and \ref{fig:SIM-Shape-1} it is shown
the deformed shape for two imposed displacements $u_{0}$, obtained
via the numerical simulations performed to solve the considered equilibrium
of the constrained micromorphic model. We remark that the following
observations can be easily inferred from the performed numerical simulations:
  
\begin{itemize}
 
\item The macroscopic shape of the specimen is completely recovered.
\item The microscopic pattern for the warp and weft is caught to a big extent.
Indeed, the real behavior of the thick yarns is almost perfectly recovered,
while, the thin yarns which are experimentally seen to undergo non-negligible
slipping, are seen to be fictively elongated. We can affirm that where
the thin yarns are seen to be elongated with respect to their initial
length a slipping is taking place.
\item The shear parameter is chosen to correctly fit the experimental load-displacement
curve. We highlight the fact that the ``force'' of the introduced
constrained micromorphic model is, to our understanding, the correct
one to be compared to the experimental one. In fact, in our micromorphic
model, we defined ``force'' the quantity which is dual to the virtual
displacement $\delta u$. From a direct observation of the irreducible
form of the work of internal forces (\ref{eq:P2_int-1}), we can infer
that the internal actions which balance a boundary externally applied
force are directly determined by equivalent elongations, shear angle
variations and microstructure-related deformation mechanisms (through
the Lagrange multiplier $\Lambda$). Analogously, we can infer that
the boundary internal actions (double-forces) which are dual to the
angle variation $\delta\varphi$ are directly determined by the localized
bending strains which occur inside the considered specimen. We state
that the double-forces which are generated by the imposition of the
angle between warp and weft at the two extremities of the specimen
do not contribute to the ``force'' which is measured by the testing
machine. Indeed, such angle is maintained fixed by the clamp device
used during the BET and a supplementary measurement tool would be
needed in order to measure such sort of couple which is generated
by the clamp in order to keep the angle constant at the boundary during
the test. In summary, we are reasonably assuming that the machine
only senses those macro and micro deformations modes which induce
macroscopic displacements (elongations, angle variations and part
of the local bending), while the remaining part of the local bending
energy which is localized at the micro-level is not sensed by the
used machine.  

\end{itemize}
 
\medskip{}
As already pointed out, the used micromorphic model possesses a feature
that the first gradient models do not posses. In particular, the reaction
at the clamps is not limited to a force, but a double-force (a couple)
arises which is needed to keep the angle between warp and weft constant
during the test. Following the methods described in subsection \ref{sub:Evaluation-of-reaction},
once calculated the solution $\left(u^{*},\varphi^{*},\Lambda^{*}\right)$
for a given imposed displacement $u_{0}$, it is possible to evaluate
the reaction double-force $\mathcal{T}$ with respect to an orthonormal
reference frame $\left\{ e_{1},e_{2}\right\} $ chosen as in Fig.
\ref{fig:Ref_Frame} by using equation (\ref{eq:ReacMicrom2}), where
we chose particular test functions $\delta\bar{u}$, $\delta\bar{\varphi}$
and $\delta\bar{\Lambda}$ such that
  
\begin{enumerate}
 
\item $\ \delta\bar{u}\cdot e_{1}=0,\qquad\forall(X,Y)\in B$ 
\item $\delta\bar{u}\cdot e_{2}=0,\qquad\forall(X,Y)\in B$ 
\item $\delta\bar{\varphi}=\frac{X}{L},\qquad\forall(X,Y)\in B$ 
\item $\delta\bar{\Lambda}=0,\qquad\forall(X,Y)\in B$.  

\end{enumerate}
 
\begin{figure}[h]
\begin{centering}
\includegraphics[scale=0.5]{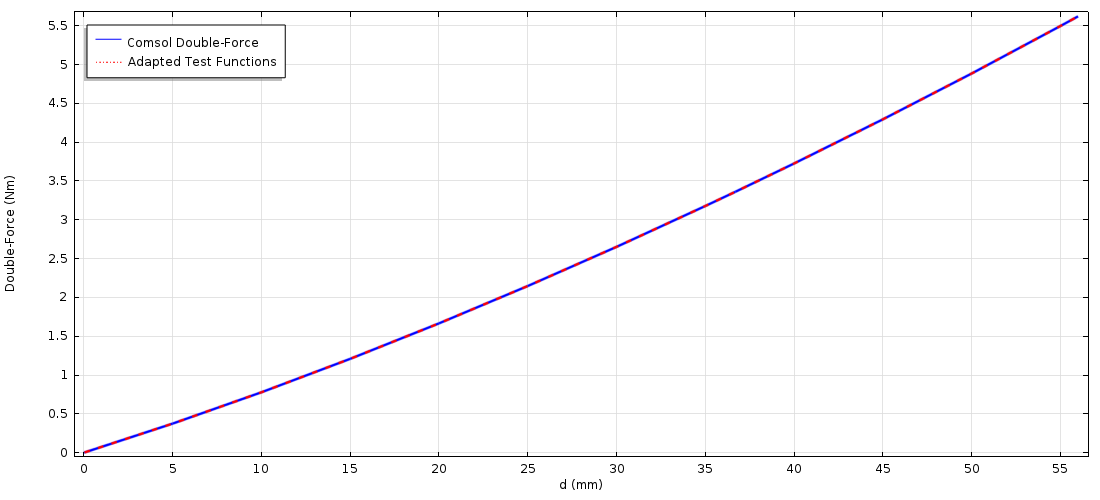}
\par\end{centering}

\protect\caption{\label{fig:SIM-LD-1-1}Double-force/displacement curve as reaction
$\mathcal{T}$ dual of $\delta\varphi$.}
\end{figure}

The results obtained are shown in figure \ref{fig:SIM-LD-1-1} where
the computed reaction double-force is also compared with the one automatically
evaluated by COMSOL$^{\circledR}$ .

\medskip{}
In a material in which the shear is the main deformation mode one
of the most important features to check is the angle between the fibers. 

\begin{figure}[H]
\begin{centering}
\includegraphics[scale=0.6]{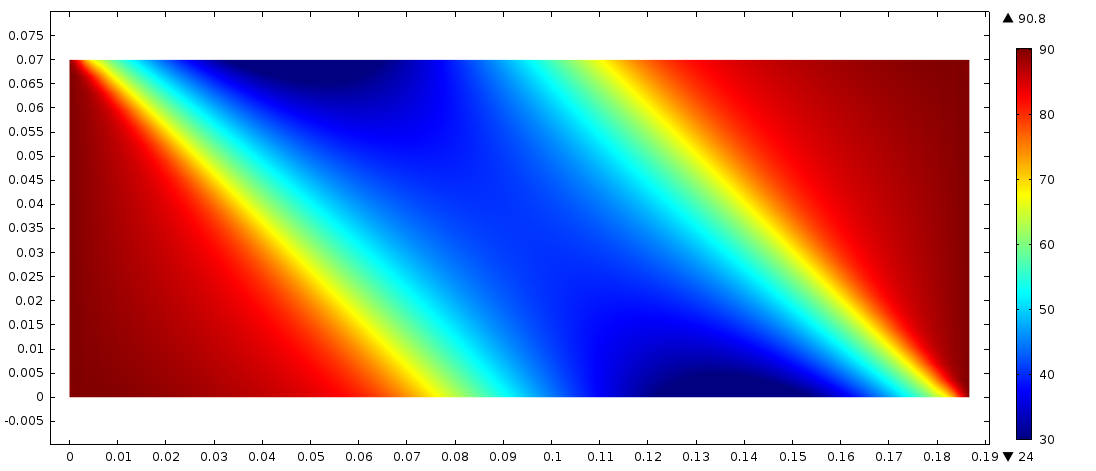}
\par\end{centering}

\protect\caption{ \label{fig:SIM-AngleVariation}Angles between the fibers in the constrained
micromorphic simulation for a displacement of of 56 mm}
\end{figure}
 It can be noticed from Fig. \ref{fig:SIM-AngleVariation} that no
angle variation occurs in the red regions, while an important and
almost constant angle variation occurs in the middle of the specimen.
The transition from one value of the angle to the other one is made
thanks to the creation of transition zones in which a smooth variation
occurs. By direct comparison with Fig. \ref{fig:SIM-Shape-1} it is
possible to remark that the three regions at constant shear angle
individuated in Fig. \ref{fig:SIM-AngleVariation} are determined
by the fact that the thick yarns remain substantially undeformed,
while the thin yarns are kinematically blocked in the standard triangular
zones of the BET, while bend in the central part of the specimen.
The change of direction of the thin yarns is the only one that determines
the angle variation between warp a weft.

We can finally remark that the bending of the thin yarns takes place
in well defined transition zones where smooth variation of the shear
angle occur.

\section{Discrete numerical simulations}

In this section we propose to set up a suitable discrete model in
which the motions of the single yarns are singularly taken into account.
To do so, we decide to use Euler-Bernoulli beams with different bending
rigidities which are interconnected with rotational and translational
elastic springs in order to mimic at best the real connections between
the yarns. The results obtained with such discrete model allow to
  
\begin{itemize}
 
\item explicitly account for the slipping of the yarns
\item better understand the potentialities and limitations of the constrained
micromorphic continuum model introduced in the previous section.  

\end{itemize}
 
The main limitation of the discrete model that we are going to present
can be found in the fact that the interactions between adjacent fibers
are all considered to be elastic. If this can be considered to be
reasonable to a certain extent, there are for sure some irreversible
mechanisms such as friction that cannot be precisely described here
and that should follow the methods presented in \cite{Gatouillat}
in order to be fully taken into account. Indeed, when unloading the
experimental specimen once that the BET has been performed, is not
sufficient to let the specimen return in its initial configuration.
This means that a certain part of the deformation is not elastic,
but is due to irreversible mechanisms such as friction. Nevertheless
a big amount of the imposed deformation is recovered and we can hence
suppose that the discrete model which we introduce here can be thought
to be a reasonable compromise between the complexity of the real microstructural
motions and the simplicity of the model that one wants to introduce.

\begin{figure}[h]
\begin{centering}
\includegraphics[scale=0.5]{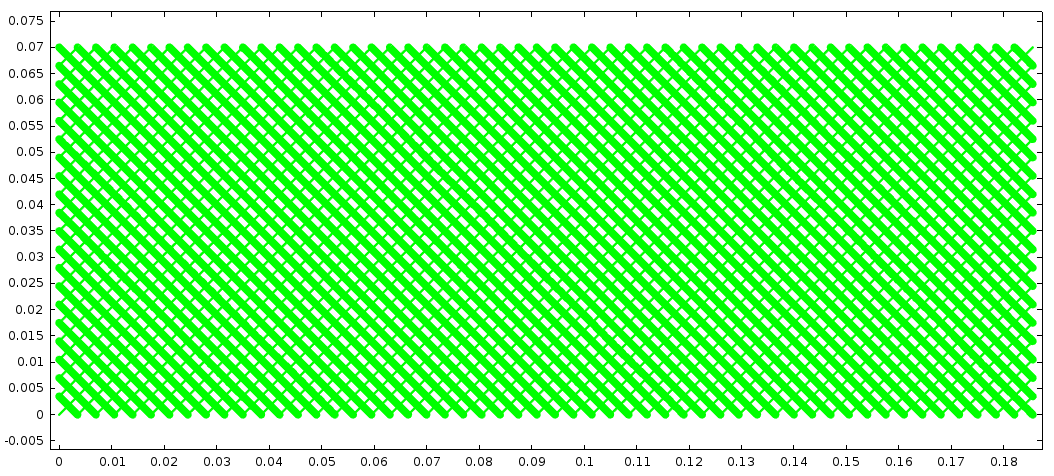}
\par\end{centering}

\protect\caption{Geometry of the discrete model: undeformed configuration.\label{fig:DIS-Geometry}}
\end{figure}

\begin{figure}[h]
\begin{centering}
\includegraphics[scale=0.35]{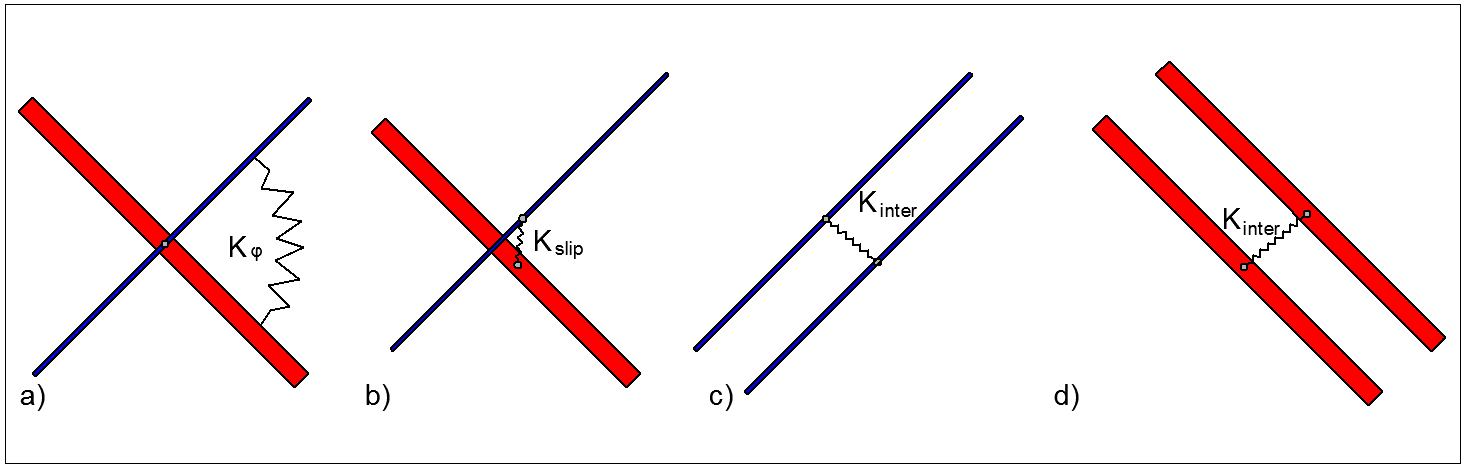}
\par\end{centering}

\protect\caption{Schematics of the elastic interconnections between warp and weft yarns.
(a) rotational spring, (b) translational spring, (c) interaction between
thin yarns and (d) interaction between thick yarns.\label{fig:DIS-Springs}}
\end{figure}

The warp and weft yarns are modeled as long Euler-Bernouilli beams
disposed at $45$ degrees with respect to the edges of the specimen
(see Fig. \ref{fig:DIS-Geometry}). The connections between the two
families of fibers is guaranteed at the contact points by rotational
and translational springs as the ones sketched in Fig. \ref{fig:DIS-Springs}.
Both beams and springs, are considered to be linear.

More particularly, the two families of yarns are modeled as beams
with axial stiffness respectively $K_{1}=EA_{1}$ and $K_{2}=EA_{2}$,
and bending stiffness respectively $K_{3}=EI_{1}$ and $K_{4}=EI_{2}$
where $E$ is the Young modulus of the material constituting the yarns,
$A_{1}$ and $A_{2}$ are the cross sections and $I_{1}$, $I_{2}$
the equivalent moments of inertia. In order to model the interactions
between the yarns the set of points in which the two families of fibers
intersect have been defined in COMSOL$^{\circledR}$. In this set
of points, the interactions between the two fibers are supposed to
be 
  
\begin{itemize}
 
\item the resistance to the variation of angle between the fibers, namely
the shear stiffness which are accounted for by the introduction of
a set of rotational springs of stiffness $K_{\varphi}$ 
\item the resistance to the slippages, described by a second set of translational
springs of stiffness $K_{slip}$ to describe the response to slippages. 
\item the mutual interactions between parallel fibers which are guaranteed
by the weaving. Such interactions are due to the presence of the orthogonal
yarns which connect the two adjacent yarns considered here. Therefore
a set of springs (modeled as trusses with axial stiffness $K_{inter}$)
has been inserted between every two couples of close interaction points
belonging to two different close yarns of the same family (see Fig
\ref{fig:DIS-Springs}).  

\end{itemize}
 
A possible downside of such discrete model is its difficult application
for bigger specimens due to the high number of degrees of freedom
needed for a proper description of the yarns response. In this optic,
continuum models are preferable to discrete ones in view of the design
of engineering structures. 

In order to compare the discrete model with the results of the continuous
ones it was implemented in $\text{COMSOL}^{\text{\ensuremath{\circledR}}}$.
In order to have a proper comparison, 

The elastic parameters used are shown in table \ref{table: Par-1}.
They have been chosen in order to be reasonably compatible with yarns
of small cross section area and Young moduli of carbon\textcolor{black}{.
The two elongation stiffnesses $K_{1}$ and $K_{2}$ have been chosen
of the same of magnitude in order to underline the fact that it is
indeed the difference in the bending stiffness of the two families
of yarns which drives the asymmetry of the macroscopic deformation.
It must be pointed out that as far as this value is high enough it
does not have a big influence in the results in terms of both displacement
and reactions. The parameters relative to the bending stiffness, the
slipping and the interaction between two fibers of the same set were
chosen via} a fit of the experimental shape of the specimen. In particular
the following characteristics were used to fit the different parameters:
the width of the specimen in the central part, the macroscopic S-deformation,
the slipping of the fibers and the distance between the fibers of
the same set. The shear stiffness $K_{\varphi}$, was chosen in order
to fit the experimental force with the reaction evaluated with the
simulations.

\begin{table}[H]
\begin{centering}
\begin{tabular}{|c|c|c|c|c|c|c|}
\hline 
$K_{1}$ & $K_{2}$ & $K_{3}$  & $K_{4}$ & $K_{\varphi}$ & $K_{slip}$ & $K_{inter}$\tabularnewline
\hline 
\hline 
50000 N & 50000 N & $0.4\ $N$\cdot$m$^{2}$ & $10^{-3}$ N$\cdot$m$^{2}$ & 2.5$10^{-4}$N$\cdot$m & 11 N/m & 11 N\tabularnewline
\hline 
\end{tabular}
\par\end{centering}

\protect\caption{Parameters of the discrete model \label{table: Par-1}}
\end{table}

\begin{figure}[h]
\begin{centering}
\includegraphics[scale=0.4]{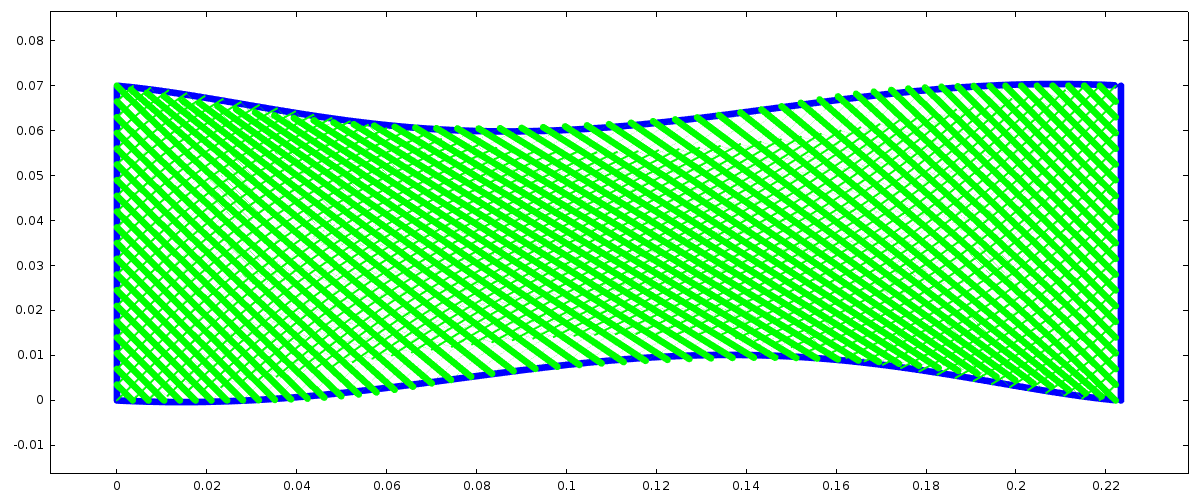}
\par\end{centering}

\protect\caption{Deformed shape in the simulation (green) and experimental (blue) for
a displacement of 37 mm \label{fig:SIM-Shape-2}}
\end{figure}

\begin{figure}[h]
\begin{centering}
\includegraphics[scale=0.4]{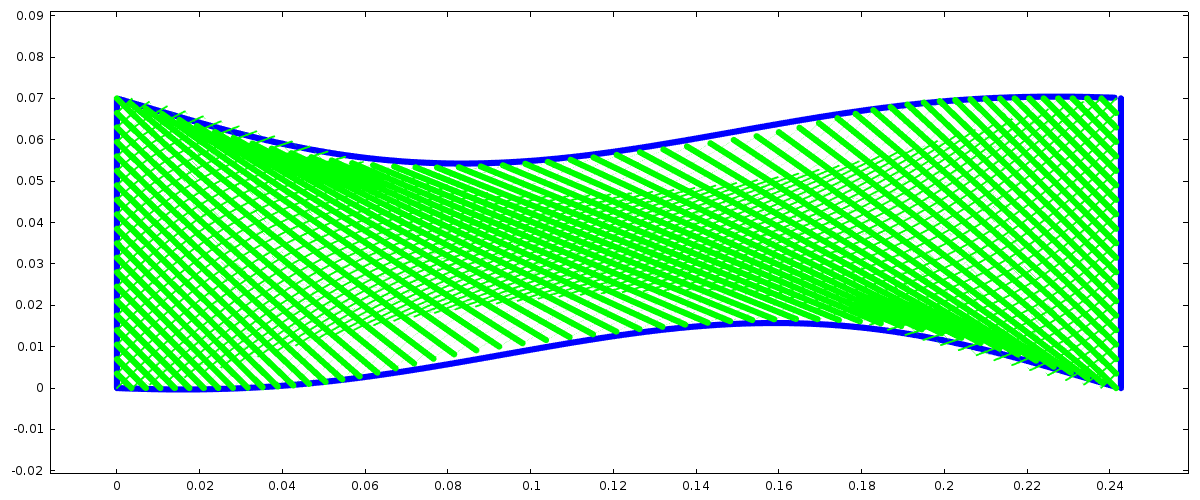}
\par\end{centering}

\protect\caption{Deformed shape in the simulation (green) and experimental (blue) for
a displacement of 56 mm \label{fig:SIM-Shape-1-1}}
\end{figure}

As done in the continuum model, in the figures \ref{fig:SIM-Shape-2}
and \ref{fig:SIM-Shape-1-1} it is possible to see that the discrete
model well describes the S-response even at different values of displacement. 

\begin{figure}[H]
\begin{centering}
\includegraphics[scale=0.4]{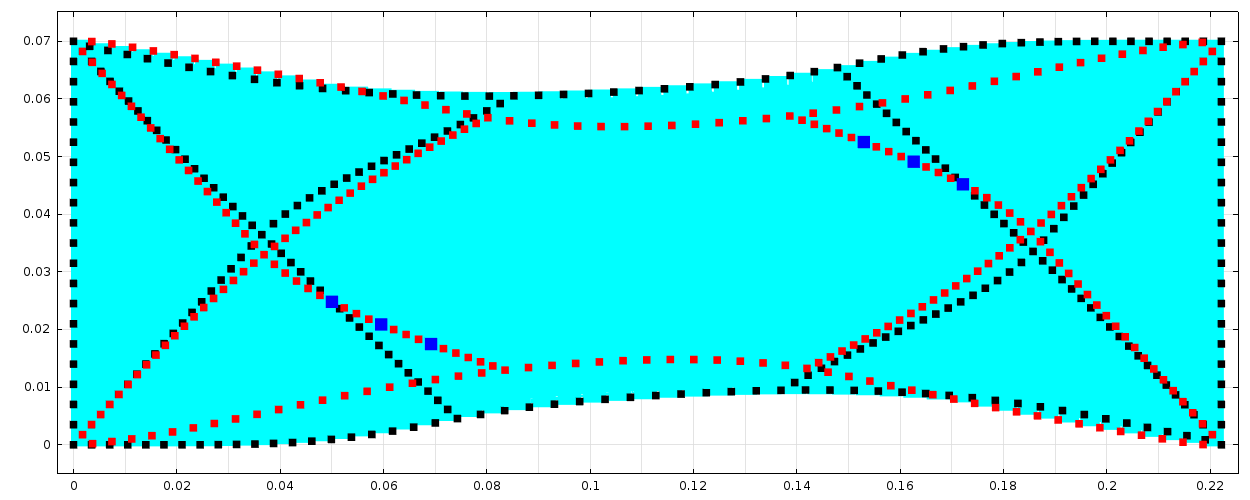}
\par\end{centering}

\protect\caption{Deformed shape for a displacement of 56 mm and some initially superimposed
points belonging to thick yarns (black) and to thin yarns (red). The
highlighted blue points correspond to the experimental ones that can
be observed in Fig. \ref{fig:sim_slip_Exp}. \label{fig:SIM-slip}}
\end{figure}

\begin{figure}[H]
\centering{}\includegraphics[width=13cm]{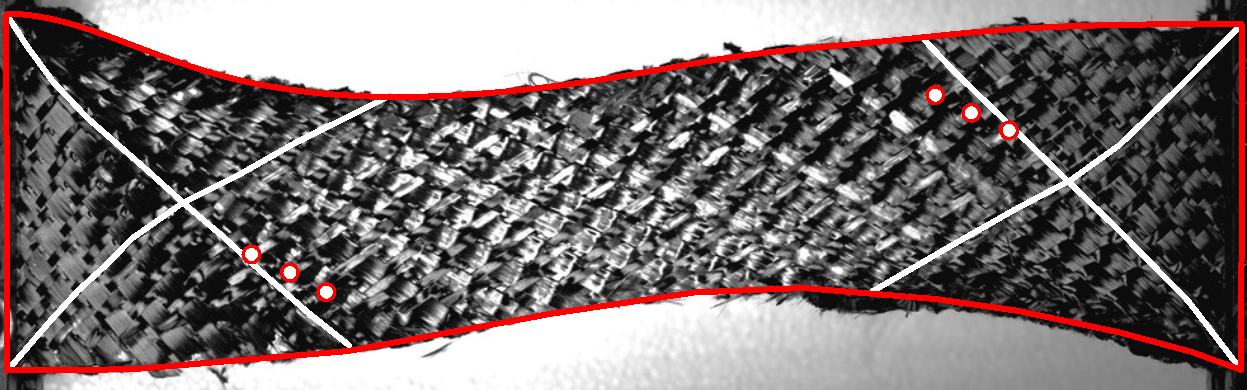}\protect\caption{\textcolor{red}{\label{fig:sim_slip_Exp}}Deformed experimental shape:
individuation of the slipping points.}
\end{figure}

 The fact that the introduced model is able to well describe the slipping
between the yarns can be more precisely seen with reference to figures
\ref{fig:SIM-slip} and \ref{fig:sim_slip_Exp}.

 In fact, the slippage of thin fibers can be recognized since 
  
\begin{itemize}
 
\item the points of the thin fibers (red) which are initially located at
the free boundaries are finally located inside the specimen. 
\item the blue points highlighted in Fig. \ref{fig:SIM-slip} and belonging
to the thin yarns are easily recognizable in the experimental deformed
specimen shown in Fig. \ref{fig:sim_slip_Exp} and it can be recognized
that the simulation well describes the experimental behavior, at least
qualitatively.  

\end{itemize}
 
The response of this model can be compared to the one obtained with
the constrained micromorphic continuum model. Indeed the different
bending stiffnesses of the two families of yarns lead to a set of
straight thick fibers and a set of strongly bent thin fibers like
in the continuum case. In this discrete simulation, nevertheless,
the slippages of the yarns are more precisely described.

\begin{figure}[H]
\begin{centering}
\includegraphics[scale=0.5]{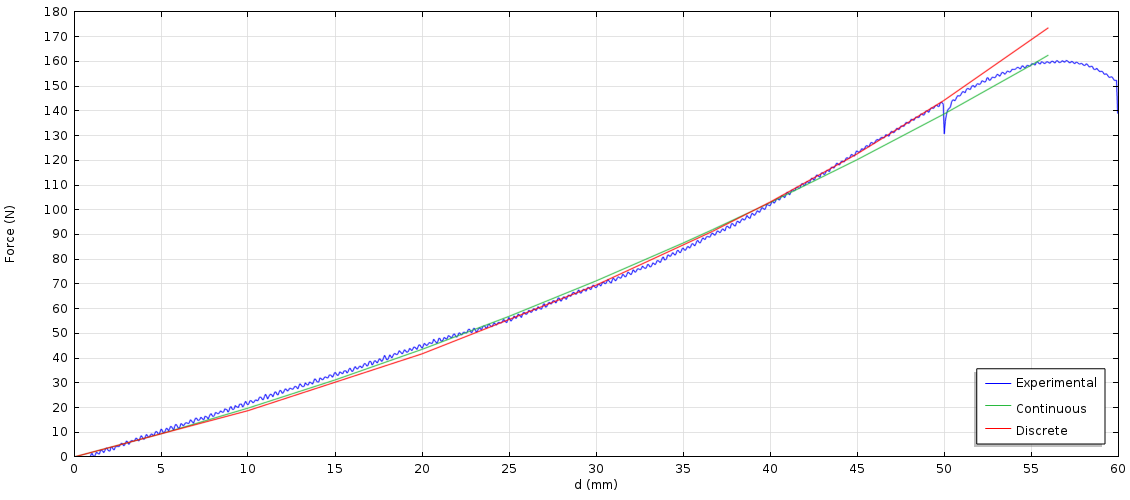}
\par\end{centering}

\protect\caption{Load-displacement curve for the experimental, discrete and second
gradient models\label{fig:DIS-LD}}
\end{figure}

\medskip{}
We show in Fig. \ref{fig:DIS-LD} that, with the constitutive choice
of the parameters shown in table \ref{table: Par-1}, also the load-displacement
curve of the discrete model results to be consistent with the experimentally
obtained one, as well as with that obtained by means of the constrained
micromorphic continuum model. We can notice that a slightly better
fitting is recovered for the discrete model at moderate strains. A
better fitting of the continuum model xould be obtained for such moderate
strains if more complex hyperelastic laws would be introduced, but
this falls outside the scope of the present paper.

\begin{figure}[H]
\begin{centering}
\includegraphics[scale=0.6]{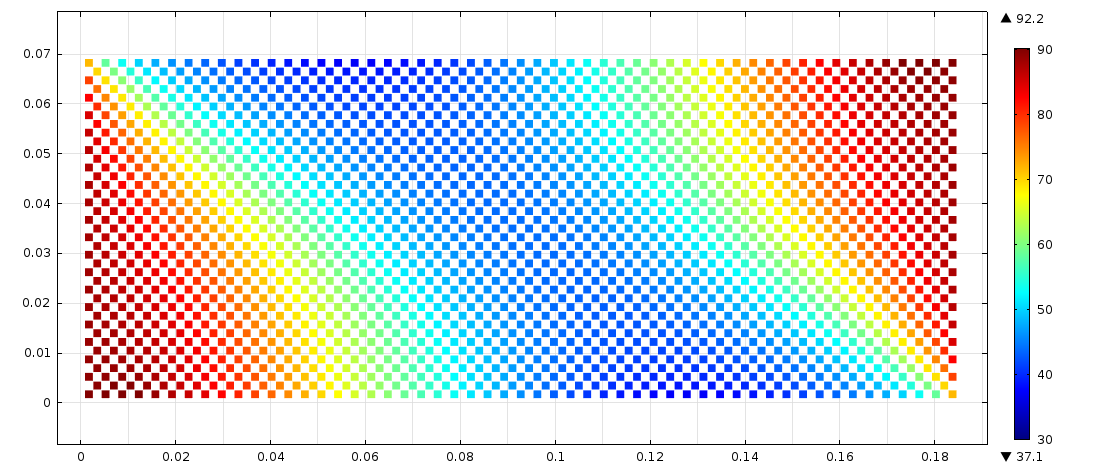}
\par\end{centering}

\protect\caption{Angles between the fibers in the discrete simulations for a displacement
of of 56 mm \label{fig:SIM-AngleVariation-1}}
\end{figure}

Finally, Fig. \ref{fig:SIM-AngleVariation-1} shows the angle between
warp and weft yarns as obtained with the discrete model., even in
this model the angle between the set of fibers is almost a constant
along the strong fibers direction while drastically changes along
the think yarns. The results of both the discrete and constrained
micromorphic continuum models, even with their very different natures,
present the same qualitative description of the experimental behavior
reconfirming the good analogy between the bending stiffness of the
yarns and the second gradient effects in the continuum. This is another
hint toward the importance of the insertion of a second gradient energy
in a continuous models in order to fully describe the phenomenological
mechanical response of the woven fabrics.

Further investigations should be more systematically directed towards
the quantitative analysis needed to precisely identify the macroscopic
constitutive parameters in terms of the microscopic properties of
the yarns. Suitable multi-scale methods as the one introduced in \cite{David7}
may be generalized to be applied to the present case. Moreover, the
description of the considered system at the microscopic scale may
take advantage of some of the results proposed in \cite{David1,David2,Rinaldi1,Rinaldi2,Rinaldi3,Rinaldi4,Rinaldi5,David3}.

\section{Conclusions}

In this paper a continuum constrained micromorphic model and a discrete
model are introduced to reproduce the Bias Extension Test on unbalanced
fabrics. We show that both models are able to account for the main
macroscopic and microscopic deformation mechanisms that take place
during the BET up to moderate strains, namely
  
\begin{itemize}
 
\item the angle variation between warp and weft tows,
\item the different bending stiffnesses of the two families of yarns (which
is at the origin of the asymmetric macroscopic S-shape of the specimen)
\item the relative slipping of the yarns.  

\end{itemize}
 
The results obtained with the two models are satisfactory when considering
small and moderate deformations so that it is conceivable to use the
proposed models on more extended experimental campaigns. This would
allow a more precise identification of the introduced constitutive
parameters, above all for what concerns the different bending stiffnesses
which are the main characteristics of fibrous composite interlocks.

Further studies should be focused on the improvement of the proposed
models in the optic of more precisely describe irreversible phenomena
as friction which can have a non-negligible role during the deformation
of woven reinforcements. 

Moreover, more extensive experimental campaigns should be carried
out in order to determine which is the strain threshold until which
the integrity of the material is preserved and both the continuum
and discrete models can be considered to be predictive. In fact, after
a given macroscopic deformation, some yarns start to be pulled out
from the specimen, so that further modeling efforts intrinsically
loose their interest.

Finally, the theoretical and numerical tools presented in this paper
can be extended in order to treat equilibrium problems concerning
3D composite interlocks. For example, the 3 point bending test on
unbalanced fabrics could be a useful way of testing and indirectly
fitting new constitutive parameters related to the out-of plane different
bending stiffnesses of the two families of yarns.

\section*{Acknowledgements }

The authors thank CNRS for the PEPS project which assured financial
support to research presented in this paper.

\pagebreak{}

\section*{Appendix A: Alternative numerical implementation of the constrained
micromorphic model: penalty method}

\begin{figure}[H]
\begin{centering}
\includegraphics[scale=0.55]{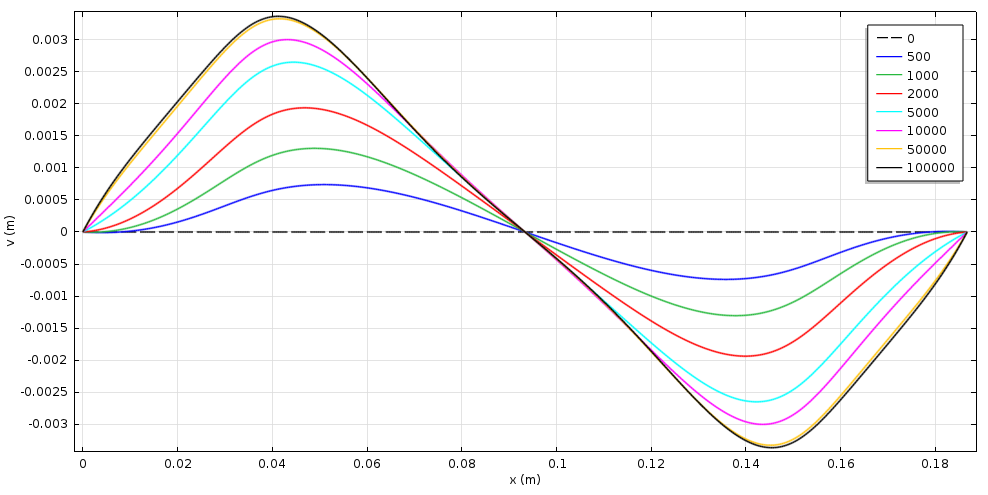}
\par\end{centering}

\protect\caption{\label{fig:SIM-Kmicro}Vertical displacement of the mean axis for
a displacement of 56 mm and different values of $K$}
\end{figure}

In this subsection, we briefly mention a method that can be used in
order to numerically implement a constrained micromorphic model as
an alternative to the method of Lagrange multipliers described in
subsection \ref{sub:Power-of-internal_Lag_mult}. It is known as ``penalty
method'' and consists in implementing a strain energy density which
takes the form
\begin{equation}
W(i_{4},i_{6},i_{8},\varphi,\nabla\varphi)=W_{I}(i_{4},i_{6},i_{8})+W_{II}(\nabla\varphi)+W_{coupling}(i_{8},\varphi),\label{eq:W_additive-1-1}
\end{equation}
where $W_{I}$ and $W_{II}$ are given in Eqs. \ref{eq:W_I} and \ref{eq:W_II}
respectively, while the coupling energy takes the form 
\[
W_{coupling}(i_{8},\varphi)=\frac{K}{2}(\varphi-i_{8})^{2},
\]
where $K$ is a constant that may ideally tend to infinity. Indeed,
in order to guarantee the boundedness of the strain energy density,
it follows that $\varphi$ must necessarily tend to $i_{8}$. We numerically
implemented such penalty method in order to test the correct convergence
of our equilibrium problem formulated with the Lagrange multipliers.
Being $K$ constant, the considered virtual variations are only $\delta u$
and $\delta\varphi$ and, moreover, the constant $K$ must be chosen
sufficiently large in order to guarantee numerical convergence of
the solution. This last feature can be easily tested by controlling
that the solution does not change when increasing the value of $K$
(see Fig. \ref{fig:SIM-Kmicro}). The constitutive parameters remain
the same as the ones used in the numerical simulation with the Lagrange
multiplier (see table \ref{table: Par}). It is possible to notice
that, suitably increasing the value of $K$ the model converges and
the limit corresponds to the solution obtained with constrained micromorphic
simulation with Lagrange multipliers (see also Fig. \ref{fig:SIM-alpha1}).

This penalty method can be seen as a useful tool for the easy implementation
of constrained micromorphic models due to their high numerical stability.


\begin{thebibliography}{10}
\bibitem{Aifantis}Aifantis E.C., 1992. On the role of gradients in
the localization of deformation and fracture. International Journal
of Engineering Science 30:10, 1279-1299

\bibitem{Aimene}Aimène Y., Vidal-Sallé E., Hagège B., Sidoroff F.,
Boisse P., 2010. A hyperelastic approach for composite reinforcement
large deformation analysis. J. Compos. Mater., 44:1, 5-26

\bibitem{Alibert}Alibert J.-J., Seppecher P., dell'Isola F., 2003.
Truss modular beams with deformation energy depending on higher displacement
gradients. Math. Mech. Solids 8:1, 51-73 

\bibitem{David1}Atai, A.A., Steigmann, D.J. (1997). On the nonlinear
mechanics of discrete networks Archive of Applied Mechanics, 67:5,
303-319

\bibitem{Neff1}Balzani D., Neff P., Schröder J., Holzapfel G.A.,
(2006). A polyconvex framework for soft biological tissues, Adjustment
to experimental data. Int. J. Solids Struct., 43, 6052-6070

\bibitem{Bleustein}Bleustein J.L., 1967. A note on the boundary conditions
of Toupin's strain gradient-theory. Int. J. Solids Structures, 3,
1053-1057.

\bibitem{Boheler}Boehler, J.P., 1987. Introduction to the invariant
formulation of anisotropic constitutive equations. In: Boehler, J.P.
(Ed.), Applications of Tensor Functions in Solid Mechanics CISM Course
No. 292. Springer-Verlag.

\bibitem{Boheler1}Boehler JP. Lois de comportement anisotrope des
milieux continus. J Méc 1978;17:153-70

\bibitem{Boisse shaping}Boisse P., Cherouat A., Gelin J.C., Sabhi
H., 1995. Experimental Study and Finite Element Simulation Of Glass
Fiber Fabric Shaping Process. Polymer Composites 16:1, 83-95

\bibitem{Cao Phi}Cao J., Akkerman R., Boisse P., Chen J., et al.,
2008. Characterization of mechanical behavior of woven fabrics: experimental
methods and benchmark results. Compos. Part A: Appl. Sci. Manuf. 39,
1037-53.

\bibitem{Casal}Casal P., 1972. La théorie du second gradient et la
capillarité. C.R. Acad. Sci. Paris, Ser. A 274, 1571-1574

\bibitem{Phil1}Charmetant A., Vidal-Sallé E., Boisse P. (2011). Hyperelastic
modelling for mesoscopic analyses of composite reinforcements. Composites
Science and Technology, 71,1623-1631

\bibitem{Phil3D}Charmetant A., Orliac J.G.,Vidal-Sallé E., Boisse
P. (2012). Hyperelastic model for large deformation analyses of 3D
interlock composite preforms. Composites Science and Technology, 72,
1352-1360

\bibitem{Cosserat}Cosserat E., Cosserat F., 1909. Théorie de Corps
déformables. Librairie Scientifique A. Hermann et fils, Paris

\bibitem{Cuomo}Cuomo M., Fagone M, 2009. Finite deformation non-isotropic
elasto-plasticity with evolving structural tensors. A framework. Il
Nuovo Cimento, 32 C:1, 55-72.

\bibitem{de Gennes}deGennes, P.G., 1981. Some effects of long range
forces on interfacial phenomena. J. Phys. Lett. 42, L377\textendash L37

\bibitem{dell_Steigmann}F. dell'Isola and D. Steigmann \textquotedblleft A
two-dimensional gradient-elasticity theory for woven fabrics\textquotedblright ,
Journal of Elasticity, vol. 118 (1), 2015, pp. 113-125

\bibitem{NthGrad} dell'Isola, F, Seppecher, P, and Madeo, A. How
contact interactions may depend on the shape of Cauchy cuts in N-th
gradient continua: approach à la D\textquoteright Alembert. ZAMP 2012;
63: 1119\textendash 1141.

\bibitem{BC2grad1}dell'Isola F., Seppecher P., 1995. The relationship
between edge contact forces, double force and interstitial working
allowed by the principle of virtual power, C.R. Acad. Sci. II, Mec.
Phys. Chim. Astron. 321, 303-308

\bibitem{BC2grad2}dell'Isola F., Seppecher P., 1997. Edge contact
forces and quasi-balanced power, Meccanica 32, 33-52 

\bibitem{Hooke2grad}dell'Isola F., Sciarra G., and Vidoli S., 2009.
Generalized Hooke\textquoteright s law for isotropic second gradient
materials, Proc. R. Soc. Lond. A 465, 2177\textendash 2196.

\bibitem{BC2grad3}dell'Isola F., Seppecher P., Madeo A., 2012. How
contact interactions may depend on the shape of Cauchy cuts in N-th
gradient continua: approach \textquotedblleft à la D\textquoteright Alembert\textquotedblright .
ZAMP, 63:6, 1119-1141

\bibitem{FdI cap1}dell'Isola F., Gouin H., Seppecher P., 1995. Radius
and surface tension of microscopic bubbles by second gradient theory.
C.R. Acad. Sci. II, Mec. 320, 211-216

\bibitem{IsolaRotoli}dell'Isola F., Rotoli G., 1995. Validity of
Laplace formula and dependence of surface tension on curvature in
second gradient fluids. Mechanics Research Communications, 22, 485-490

\bibitem{FdI cap2}dell'Isola F., Gouin H., Rotoli G., 1996. Nucleation
of Spherical shell-like interfaces by second gradient theory: numerical
simulations, Eur. J. Mech. B, Fluids 15:4, 545-568

\bibitem{FdIGuarHutter}dell'Isola F., Guarascio M., Hutter K., 2000.
A variational approach for the deformation of a saturated porous solid.
A second-gradient theory extending Terzaghi\textquoteright s effective
stress principle. Archive of Applied Mechanics. 70, 323-337.

\bibitem{FdIwaves}dell'Isola F., Madeo A., Placidi L., 2012. Linear
plane wave propagation and normal transmission and reflection at discontinuity
surfaces in second gradient 3D Continua. Zeitschrift fur Angewandte
Mathematik und Mechanik (ZAMM), 92:1, 52-71

\bibitem{Yves1}Dumont J.P., Ladeveze P., Poss M., Remond Y., 1987.
Damage mechanics for 3-D composites Composite structures, 8:2, 119-141

\bibitem{Victor00}Eremeyev V. A., Lebedev L. P., Altenbach H. (2013).
Foundations of micropolar mechanics. Springer, Heidelberg.

\bibitem{Victor2}Eremeyev V.A., 2005. Acceleration waves in micropolar
elastic media. Doklady Physics 50:4, 204-206

\bibitem{EringenBook}Eringen A. C., 2001. Microcontinuum field theories.
Springer-Verlag, New York.

\bibitem{MadeoBias}Ferretti M., Madeo A., dell'Isola F., Boisse P.
2014. Modeling the onset of shear boundary layers in fibrous composite
reinforcements by second-gradient theory, Zeitschrift für angewandte
Mathematik und Physik, Volume 65, Issue 3 , pp 587-612 

\bibitem{Forest0}Forest, S., Sievert, R. 2006. Nonlinear microstrain
theories. Int. J. Solids Struct., 43, 7224-7245.

\bibitem{Forest}Forest S. 2009. Micromorphic Approach for Gradient
Elasticity, Viscoplasticity, and Damage. Journal of Engineering Mechanics,
135:3, 117-131.

\bibitem{Forest Aifantis}Forest S., Aifantis E.C., 2010. Some links
between recent gradient thermo-elasto-plasticity theories and the
thermomechanics of generalized continua. Int. J. Solids. Struct. 47:(25-26),
3367-3376

\bibitem{Gatouillat} Gatouillat S., Bareggi A., Vidal-Sallé E., Boisse
P., 2013. Meso modelling for composite preform shaping \textendash{}
Simulation of the loss of cohesion of the woven fibre network. Composites
Part A, vol. 54, pp. 135-144.

\bibitem{Germain}Germain, P., 1973. La méthode des puissances virtuelles
en mécanique des milieux continus. Première partie. Théorie du second
gradient. J. Mécanique 12, 235-274.

\bibitem{Germain2}Germain, P., 1973. The method of virtual power
in continuum mechanics. Part 2: Microstructure. SIAM J. Appl. Math.
25, 556-575

\bibitem{NeffMicromE}Ghiba ID, Neff P, Madeo A, Placidi L, Rosi G.
(2013) The relaxed linear micromorphic continuum: existence, uniqueness
and continuous dependence in dynamics. Mathematics and Mechanics of
Solids, 1081286513516972

\bibitem{GreenRivlin}Green A.E., Rivlin R.S., 1964. Multipolar continuum
mechanics. Archive for Rational Mechanics and Analysis, 17: 2, 113-147

\bibitem{Harrison}Harrison P., Clifford M.J., Long A.C. 2004. Shear
characterisation of viscous woven textile composites: a comparison
between picture frame and bias extension experiments. Composites Science
and Technology, 64, 1453-1465

\bibitem{David2}Haseganu, E.M., Steigmann, D.J. (1996). Equilibrium
analysis of finitely deformed elastic networks. Computational Mechanics,
17:6, 359-373

\bibitem{Holzapfel}Holzapfel, G.A., Gasser, T.C., Ogden, R.W., 2000.
A new constitutive framework for arterial wall mechanics and a comparative
study of material models. Journal of Elasticity 61, 1-48.

\bibitem{Holzapfel book}Holzapfel, G.A., 2000. Nonlinear Solid Mechanics,
Wiley.

\bibitem{Itskov0}Itskov M., 2007. Tensor Algebra and Tensor Analysis
for Engineers (Springer) 

\bibitem{Itskov}Itskov M., Aksel N., 2004. A class of orthotropic
and transversely isotropic hyperelastic constitutive models based
on a polyconvex strain energy function. Int. J. Solids Struct., 41,
3833\textendash 3848

\bibitem{Itskov 1}Itskov M. 2000. On the theory of fourth-order tensors
and their applications in computational mechanics. Comput Methods
Appl. Mech. Eng., 189:2, 419-38

\bibitem{NeffMicromC}Klawonn A, Neff P, Rheinbach O, Vanis S., 2011.
FETI-DP domain decomposition methods for elasticity with structural
changes: P-elasticity. ESAIM: Mathematical Modelling and Numerical
Analysis 45 (03), 563-602

\bibitem{Lee e Phil}Lee W., Padvoiskis J., Cao J., de Luycker E.,
Boisse P., Morestin F., Chen J., Sherwood J., 2008. Bias-extension
of woven composite fabrics. Int J Mater Form Suppl 1:895-898

\bibitem{madeo porous}Madeo A., dell'Isola F., Ianiro N., SciarraG.,
2008. A variational deduction of second gradient poroelasticity II:
An application to the consolidation problem, J. Mech. Mater. Struct.
3:4, 607-625

\bibitem{madeoWavePor}Madeo A., Djeran-Maigre I., Rosi G., Silvani
C., 2012. The Effect of Fluid Streams in Porous Media on Acoustic
Compression Wave Propagation, Transmission and Reflection. Continuum
Mechanics and Thermodynamics, DOI: 007/s00161-012-0236-y

\bibitem{incomplete}Madeo A., Ghiba I.D., Neff P., Münch, I., 2015,
Incomplete traction boundary conditions in the Grioli-Koiter-Mindlin-Toupin
indeterminate couple stress model. arXiv:1505.00995

\bibitem{Yves3}Makradi A., Ahzi S., Garmestani H., Li D.S., Rémond
Y., 2010. Statistical continuum theory for the effective conductivity
of fiber filled polymer composites: Effect of orientation distribution
and aspect ratio A Mikdam. Composites Science and Technology 70 :3,
510-517 

\bibitem{Yves4}Mikdam A., Makradi A., Ahzi S., Garmestani H., Li
D.S., Rémond Y., 2009. Effective conductivity in isotropic heterogeneous
media using a strong-contrast statistical continuum theory. Journal
of the Mechanics and Physics of Solids 57:1, 76-86

\bibitem{Marsden} J.E. Marsden, T.J.R Hughes, 1994. Mathematical
Foundations of Elasticity. Dover Publications inc., New York.

\bibitem{Mindlin}Mindlin R.D., 1964. Micro-structure in linear elasticity.
Archs ration. Mech. Analysis 51-78

\bibitem{Mindlin1}Mindlin, R.D., Eshel, N.N., 1968. On first strain-gradient
theories in linear elasticity. Int. J. Solids Struct. 4, 109-124. 

\bibitem{David7}Nadler, B., Papadopoulos, P., Steigmann, D.J. (2006).
Multiscale constitutive modeling and numerical simulation of fabric
material. International Journal of Solids and Structures, 43 (2),
pp. 206-221

\bibitem{David4}Nadler, B., Steigmann, D.J. (2003). A model for frictional
slip in woven fabrics. Comptes Rendus - Mecanique, 331 (12), pp. 797-804

\bibitem{Neff1-Microm}Neff P, Forest S. 2007. A geometrically exact
micromorphic model for elastic metallic foams accounting for affine
microstructure. Modelling, existence of minimizers, identification
of moduli and computational results. Journal of Elasticity 87 (2-3),
239-276

\bibitem{Neff2-Microm} Neff P. (2006) Existence of minimizers for
a finite-strain micromorphic elastic solid. Proceedings of the Royal
Society of Edinburgh: Section A Mathematics 136, 997-1012 

\bibitem{Neff3-Microm}Neff P. (2014) Existence of minimizers in nonlinear
elastostatics of micromorphic solids. Encyclopedia of Thermal Stresses,
1475-1485 

\bibitem{NeffMicromA}Neff P, Ghiba ID, Madeo A, Placidi L, Rosi G.
(2013) A unifying perspective: the relaxed linear micromorphic continuum.
Continuum Mechanics and Thermodynamics 26 (5), 639-681

\bibitem{NeffMicromB}Neff P, Jeong J, Münch I, Ramezani H. (2008)
Mean field modeling of isotropic random Cauchy elasticity versus microstretch
elasticity. Zeitschrift für angewandte Mathematik und Physik 60 (3),
479-497

\bibitem{NeffMicromD} Neff P, Jeong J, Ramézani H. (2009) Subgrid
interaction and micro-randomness\textendash Novel invariance requirements
in infinitesimal gradient elasticity. International Journal of Solids
and Structures 46 (25), 4261-4276

\bibitem{NeffMicromF}Neff P., Ghiba ID, Lazar M, Madeo A., (2014)
The relaxed linear micromorphic continuum: well-posedness of the static
problem and relations to the gauge theory of dislocations. The Quarterly
Journal of Mechanics and Applied Mathematics 

\bibitem{Ogden}Ogden R.W. Non-linear elastic deformations. New York:
Wiley and Sons; 1984. 

\bibitem{Ogden CISM}Ogden R.W., 2003. Nonlinear Elasticity, Anisotropy,
Material Stability and Residual stresses in Soft Tissue. CISM Courses
and Lectures Series 441, 65-108

\bibitem{Yves2}Oshmyan V.G., Patlazhan S.A., Rémond Y., 2006. Principles
of structural-mechanical modeling of polymers and composites. Polymer
Science Series A 48:9, 1004-1013

\bibitem{Peng}Peng X., Guo Z., Du T., Yu W.R., 2013. A Simple Anisotropic
Hyperelastic Constitutive Model for Textile Fabrics with Application
to Forming Simulation. Composites: Part B, DOI: http://dx.doi.org/10.1016/j.compositesb.2013.04.014

\bibitem{Pideri-Sepp}Pideri C., Seppecher P., 1997. A second gradient
material resulting from the homogenization of an heterogeneous linear
elastic medium. Contin. Mech. Thermodyn. 9:5, 241-257

\bibitem{Piola}Piola G., 1846. Memoria intorno alle equazioni fondamentali
del movimento di corpi qualsivogliono considerati secondo la naturale
loro forma e costituzione. Modena, Tipi del R.D. Camera

\bibitem{Placidi}Placidi L., Rosi G., Giorgio. I., Madeo A., 2013.
Reflection and transmission of plane waves at surfaces carrying material
properties and embedded in second gradient materials. Mathematics
and Mechanics of Solids, DOI: 10.1177/1081286512474016. 

\bibitem{Raoult Annie}Raoult A., 2009. Symmetry groups in nonlinear
elasticity: An exercise in vintage mathematics. Communications on
Pure and Applied Analysis 8:1, 435-456

\bibitem{Rinaldi1}Rinaldi A., Krajcinovic K., Peralta P., Lai Y.-C.,
(2008). Modeling Polycrystalline Microstructures With Lattice Models:
A Quantitative Approach. Mech. Mater., 40, 17-36 

\bibitem{Rinaldi2}Rinaldi A. (2009). A rational model for 2D Disordered
Lattices Under Uniaxial Loading. Int. J. Damage Mech., 18, 233-57

\bibitem{Rinaldi3}Rinaldi A., Lai Y.C., (2007), Damage Theory Of
2D Disordered Lattices: Energetics And Physical Foundations Of Damage
Parameter. Int. J. Plasticity, 23, 1796-1825 

\bibitem{Rinaldi4} Rinaldi A. (2011). Statistical model with two
order parameters for ductile and soft fiber bundles in nanoscience
and biomaterials. Phys Rev E , 83(4-2) 046126

\bibitem{Rinaldi5}Rinaldi A. (2013). Bottom-up modeling of damage
in heterogeneous quasi-brittle solids. Continuum Mechanics and Thermodynamics,
25, Issue 2-4, 359-373

\bibitem{Guyader}Rosi G., Madeo A., Guyader J.-L., 2013. Switch between
fast and slow Biot compression waves induced by second gradient microstructur
at material discontinuity surfaces in porous media. Int. J. Solids
Struct., 50:10, 1721-1746

\bibitem{NeffBis}Schröder J., Balzani D., Neff P., 2005. A variational
approach for materially stable anisotropic hyperelasticity. Int. J.
Solids Struct., 42, 4352-4371

\bibitem{CoussiSciarra} Sciarra G., dell'Isola F., Coussy O., 2007.
Second gradient poromechanics, Int. J. Solids Struct, 44:20, 6607-6629

\bibitem{sciarra}Sciarra G., dell' Isola F., Ianiro N., Madeo A.,
2008. A Variational Deduction of Second Gradient Poroelasticity I:
General Theory, Journal of Mechanics of Materials and Structures 3:3
507-526 

\bibitem{Seppecher Exotic}Seppecher P., Alibert J.-J., dell\textquoteright Isola
F., 2011. Linear elastic trusses leading to continua with exotic mechanical
interactions. Journal of Physics: Conference Series, 319

\bibitem{Spencer}Spencer A.J.M., Constitutive theory for strongly
anisotropic solids, in Continuum Theory of Fibre- Reinforced Composites,
CISM International Centre for Mechanical Sciences Courses and Lecture
Notes, 282, Spencer A.J. M. Ed., Springer, 1984.

\bibitem{David3}Steigmann, D.J. (1992). Equilibrium of prestressed
networks. IMA Journal of Applied Mathematics (Institute of Mathematics
and Its Applications), 48:2, 195-215

\bibitem{David6} Steigmann, D.J. (2002). Invariants of the stretch
tensors and their application to finite elasticity theory. Mathematics
and Mechanics of Solids, 7:4, 393-404

\bibitem{David5}Steigmann, D.J. (2003). Frame-invariant polyconvex
strain-energy functions for some anisotropic solids Mathematics and
Mechanics of Solids, 8:5, 497-506

\bibitem{Toupin}Toupin R. 1964. Theories of elasticity with couples-stress.
Arch. Rational Mech and Anal., 17, 85-112

\bibitem{Trianta}Triantafyllidis N., Aifantis E.C.A., 1986. Gradient
approach to localization of deformation, I. Hyperelastic materials.
J. Elast. 16:3, 225-237\end{thebibliography}
\end{document}